\documentclass[a4paper,12pt,reqno]{amsart}

\usepackage{a4}
\usepackage{verbatim}
\usepackage{comment}
\usepackage{color}
\usepackage{todonotes}
\usepackage{verbatim}

\usepackage{amssymb,stmaryrd}
\usepackage{color}
\usepackage{graphicx}
\usepackage{floatflt}
\usepackage{float} 		%damit H funktioniert bei bildern
\usepackage{amsthm}
\usepackage{amscd}
\usepackage{hyperref}

\newtheorem{definition}{Definition}

\newtheorem*{theorem*}{Theorem}
\newtheorem{lemma}[definition]{Lemma}
\newtheorem*{lemma*}{Lemma}
\newtheorem{remark}[definition]{Remark}

\newtheorem*{corollary*}{Corollary}

\def\XXint#1#2#3{{\setbox0=\hbox{$#1{#2#3}{\int}$}
		\vcenter{\hbox{$#2#3$}}\kern-.5\wd0}}

\makeatletter
\@addtoreset{definition}{section}
\@addtoreset{equation}{section}
\makeatother

\DeclareMathOperator{\Res}{Res}

\allowdisplaybreaks[4]

\title[Quantum Curve for strips from TR and open GW/DT invariants]{Quantum Curve for strip geometries, Topological Recursion and open GW/DT invariants}

\author{Sibasish Banerjee}

\address{IHES, 35 Route de Chartres, 91440 Bures-sur-Yvette, France}

\email{sbanerjee@ihes.fr}

\author{Alexander Hock}

\address{Section of Mathematics, University of Geneva, Rue du Conseil-G\'{e}n\'{e}ral 7-9,
	1205 Geneva, Switzerland} 
\email{alexander.hock@unige.ch}

\begin{document}
\maketitle

\begin{abstract}

Open topological string partition function gives rise to open Gromov-Witten invariants, open Donaldson-Thomas invariants and 3D-5D BPS indices. Utilizing the remodelling conjecture which connects topological recursion and topological string theory, in this paper we study open topological string theory for the subclass of toric Calabi-Yau threefold known as strip geometries. For this purpose, certain new developments in the theory of topological recursion are applied as its extension to Logarithmic Topological Recursion (Log-TR) and the universal $x$--$y$ duality. Through this we  derive the open topological string partition function and also the associated quantum curve. We also explain how this is related the open Donaldson-Thomas partition function associated with certain symmetric quivers, exponential networks and $q$-Barnes type integrals. In the process, we also connect how 3D-5D wall crossing affects these partition functions as one varies $x$, in examples.

\end{abstract}

\section{Introduction and summary}
\label{sec:intro}

One of the cornerstones in testing predictions of mirror symmetry is the explicit computation of Gromov–Witten (GW) invariants. In \cite{Banerjee:2025qgx}, we studied the connection between closed GW invariants and Donaldson–Thomas (DT) invariants for a particular class of toric Calabi–Yau (CY) threefolds, known as strip geometries. A simplifying feature of these geometries is the absence of compact four-cycles. With the advent of $x$–$y$ duality \cite{Alexandrov:2022ydc,Hock:2023dno,Alexandrov:2023tgl,Hock:2025sxq,Hock:2025wlm}, it has become possible to compute the topological string partition function to arbitrary order in $\hbar$ for such CY threefolds directly from topological recursion. In \cite{Banerjee:2025qgx}, we exploited this fact to compute closed GW invariants and matched them against DT invariants, obtained through various techniques in \cite{Iqbal:2004ne, Szendr_i_2008, Sulkowski:2009rw, Banerjee:2018syt, Banerjee:2019apt, Mozgovoy:2020has}. There, we also explained the importance of this comparison in the context of homological mirror symmetry and the interplay among different enumerative invariants of Calabi–Yau threefolds. In the present article, we go beyond the closed string sector. Using $x$–$y$ duality, we shed light on the relation between open GW and open DT invariants, which remain far less understood—particularly regarding their significance from the perspective of mirror symmetry. For this class of toric Calabi–Yau threefolds, we aim to uncover new aspects of this relation by applying the techniques of topological recursion.

Recall, that for a toric CY threefold $\mathcal{Y}$, in the A-model has a description in terms of the symplectic quotient. One can associate a toric graph to $\mathcal{Y}$, which for our particular case of interest (strip geometries), will be toric trees. As said before, they have no compact four cycle and are obtained by gluing $\mathcal{O}(-1)\oplus \mathcal{O}(-1) \rightarrow \mathbb{P}^1$ and $\mathcal{O}(-2)\oplus \mathcal{O} \rightarrow \mathbb{P}^1$, the two resolutions of $\mathbb{P}^1.$ The mirror CY threefolds are given by $\mathcal{X} = \{(x,y,u,v) \vert A(e^x,e^y) = uv\} \subset \mathbb{C}^4$, where $A(e^x,e^y)= 0$ is called the mirror curve. For our examples, they take the following form 
\begin{align}\label{spectralcurveintro}
    A(e^x,e^y)=(1-e^y)\prod_{j=1}^s(1-\beta_j e^y)+(-1)^f e^xe^{y(1+f)} \prod_{j=1}^r(1-\alpha_j e^y)=0.
\end{align}

{\it Our open GW/open DT correspondence for strip geometries is the following: Using Log-TR through $x$--$y$ duality, we construct a quantum curve which has the classical limit \eqref{spectralcurveintro}. The form of the quantum curve itself depends on the choice of a basepoint on the $\mathbb{C}_x$ plane. For certain choices of $x$, we show that the solutions of the associated $q$-difference equation defined by the quantum curve are of the form of \eqref{eq:DTfnintro}, from which we can read off the open DT invariants.}\\

Let us first recapitulate the computation of the open GW invariants. Starting from the CY $\mathcal{Y}$ (toric CY threefold, mirror to $\mathcal{X}$) and a toric noncompact Lagrangian $L$, such that $b_1(L)=1$, one  can consider stable maps $u : (\Sigma_{g,n},\partial\Sigma_{g,n}) \rightarrow (\mathcal{Y},L)$. The toric Lagrangians branes preserve $A$-twisted boundary conditions which provide the A-model description. This is also the setup described in \cite{Aganagic:2000gs,Aganagic:2001nx}. In \cite{Graber:2001dw}, a localization formula for GW invariants was proposed based on the findings of \cite{Katz:2001vm,Li_2006}, where in the latter two it was found that there is an algebraic moduli space where the space of stable maps $u$ embeds. We will not go to the details, but sidestep the issue using remodelling conjecture (which is now a theorem already) of \cite{Bouchard:2007ys}. Then the computation of open GW invariants is equivalent to the computation of certain periods on the mirror curve, where the differentials \eqref{eq:TR-introLog} are defined through Log-TR and the open GW partition function is given by \eqref{wavefunction}. 

Indeed, it was clarified in \cite{Aganagic:2001nx} that the choice of $L$ in the A-side is equivalent to choice of flat coordinates on the B-model side. The integrals that appear in \cite{Aganagic:2000gs,Aganagic:2001nx} are solutions to {\it extended} Picard-Fuchs equations \cite{Lerche:2002yw, Mayr:2001xk,Forbes:2005xt}. Based on remodelling conjecture, and with the help of $x$--$y$ duality, our concrete strategy of deriving the open GW invariant partition function is as follows. Through interchanging $x$ and $y$, for the class of curves pertinent for this paper, \eqref{spectralcurveintro}, one obtains a fairly simple expression for the wave function, for example \eqref{dualwavepoch}, satisfying a linear difference equation \eqref{differencey}. In section \textsection \ref{sec:TRStrip}, we describe how to obtain the quantum curve and the difference equation for the wavefunction in the original TR side. One important point is that, the form of the difference equation depends on the choice of basepoint and to solve these difference equations using Frobenius' method, one needs to choose the right coordinates in the summand. We also obtain the expressions for the solutions, in their regime of validity in section \textsection \ref{sec:TRStrip} and show that the expressions obtained are of Heine's $q$-hypergeometric type. Such expressions for this class of geometry was also obtained in \cite{Panfil:2018faz}. The novelty of our work is that, we can compute this partition function for different basepoints $x$, using TR concretely.

This form of the generating series of the open GW invariants satisfy the {\it admissibility} criteria for generating series of (open) DT invariants for this class of toric CY, \cite{kontsevich2011cohomologicalhallalgebraexponential,Efimov_2012,Kucharski:2017ogk}.  The generating function of the open  DT invariants obtained from the associated symmetric quiver is of the following form, where the quiver is encoded in the matrix $C_{ij}$ (entries $C_{ij}$ indicate the number of arrows between two vertices $i$ and $j$) 
\begin{equation}
\label{eq:DTfnintro}
\begin{split}
P_C (X_1,...,X_m) &= \sum_{d_1,...,d_m} \frac{(-q^{1/2})^{\sum_{i,j}^mC_{ij} d_id_j}}{(q;q)_{d_1}(q;q)_{d_2}...(q;q)_{d_m}} X_1^{d_1}X_2^{d_2}...X_m^{d_m}
\\ & 
= \prod_{d_1,...,d_m} \prod_{j\in \mathbb{Z}} \prod_{k=1}^\infty \bigg(1- X_1^{d_1}X_2^{d_2}...X_m^{d_m} q^{k+j(j-1)/2}\bigg)^{(-1)^{j+1} \Omega_{d_1,...,d_m;j}}, 
\end{split}
\end{equation}
where the second line defines the motivic DT invariants. We think of all $X_i=e^{x_i}$ here as formal $\mathbb{C}^\times$ variables. 

A more physical description of these open DT invariants can be given based on the difference equations that we compute using TR \cite{Banerjee:2018syt,Banerjee:2019apt,DelMonte:2024dcr}. Our open GW partition function is an asymptotic series in $\hbar$. In the process of writing the quantum curves in \eqref{eq:TRAhatbpinfty} and \eqref{eq:Ahatbp1}, we already have implicitly made certain choice for $\vartheta = \arg (\hbar)$, which we keep fixed throughout. However, varying $x$ also produces wall-crossing of open BPS states. \footnote{Compactifying M-theory on $(\mathcal{X},L)$ engineers a 5D $N=1$, gauge theory $T^{[5D]}(\mathcal{X})$ coupled to a 3D $N=2$ gauge theory $T^{[3D]}(L)$ with $x$ being the Fayet-Iliopoulos (FI) parameter. Here we are varying this FI parameter $x$ resulting change of BPS spectra, through change in the stability condition of 3D BPS states in $T^{[3D]}(L)$ . This step 1 (out of 2) of nonabelianization process described in \cite{Banerjee:2018syt}.} We will come to their discussion in slightly more details in section \textsection \ref{sec:openDT} (and we refer to \cite{Banerjee:2018syt} for more details). Once the B-brane partition functions are computed for different $x$-coordinates written as $q$-series, one can relate them via wall-crossing, which evidences towards their relation to the open DT invariants, through the change of spectrum of the 3D-5D BPS states.  

In the rest of the paper, we recapitulate Log-TR in section \textsection \ref{sec:TRStrip}. After that we describe $x$--$y$ duality to derive the dual partition functions, for the curve \eqref{spectralcurveintro}, and then we compute the quantum curve and the wave function (B-brane partition function) for the original side, for different choices of basepoints. Then in section \textsection \ref{sec:openDT}, we connect the open GW partition functions computed in the B-model side to open DT invariants. We explain this connection in two ways: firstly, we follow \cite{Panfil:2018faz} and show that (and provide examples) they can be rewritten in terms of generating series of DT invariants associated to symmetric quivers, as in \cite{kontsevich2011cohomologicalhallalgebraexponential,Efimov_2012,Kucharski:2017ogk}. We also explain how the different choices of basepoints give rise to different forms of these quiver generating series and how the corresponding open DT invariants are related by wall-crossing. We also explain the connection to exponential networks and finally show that these quiver generating series can be rewritten as $q$-Barnes integrals which illustrates various features of these generating series. We also explain how the framing transformation parameter $f\in \mathbb{Z}$ affects the partition functions.

\subsection*{Acknowledgment}
Special thanks go to Kohei Iwaki who first brought this problem to our attention. We would like to particularly thank Mauricio Romo for sharing many ideas regarding the computations. We would also like to thank Nafiz Ishtiaque, Saebyeok Jeong, Maxim Kontsevich, Pietro Longhi, Olivier Marchal, Nicolas Orantin, Raphael Senghaas, Valdo Tatitscheff, Johannes Walcher and Campbell Wheeler for discussing related problems.   We thank the organizers of the trimester program ``Higher Rank Geometric Structures" at the Institut Henri Poincaré, Paris, where the collaboration ensued. SB acknowledges the working conditions and hospitality provided by CERN and by the University of Geneva. The work of S.B.
has been supported in part by the ERC-SyG project “Recursive and Exact New Quantum
Theory” (ReNewQuantum), which received funding from the European Research Council
(ERC) under the European Union’s Horizon 2020 research and innovation program, grant
agreement No. 810573. Currently, S.B. is supported by Simons' general program. S.B. also thanks Universit\'e de Gen\`eve for providing excellent working conditions.  A.H. was funded by
	the Deutsche Forschungsgemeinschaft (DFG, German Research
	Foundation) -- project ID 551478549 and by the Swiss National Science Foundation (SNSF) through the
Ambizione project “TRuality: Topological Recursion, Duality and Applications”
under the grant agreement PZ00P2 223297.

\section{Quantum Curve from Topological Recursion for Strip Geometry}
\label{sec:TRStrip}

In this section we recall the background on topological recursion (TR) and discuss recent advances that enable us to  explicitly derive the quantum curve and the associated wave function for strip geometries. In the present work, however, these quantities will not be obtained directly through the recursive definition of TR. Instead, we will make use of a duality, namely the $x$–$y$ duality, and an extension of TR the so-called logarithmic topological recursion (Log-TR), which together provide the right way to compute those objects.

\subsection{Background on Topological Recursion}
Topological recursion is a recursive procedure which, given the initial data of a spectral curve, generates an infinite tower of multidifferentials $\omega_{g,n}$ on $n$ copies of the curve. The construction, introduced in \cite{Eynard:2007kz}, proceeds recursively in $2g+n-2$ with $g\in\mathbb{Z}_{\geq 0}$ and $n\in\mathbb{Z}_{> 0}$, and admits an elegant diagrammatic interpretation.

The input, called the spectral curve, consists of a quadruple $(\Sigma,x,y,B)$ where $\Sigma$ is a Riemann surface, $x,y:\Sigma\to\mathbb{C}$ are meromorphic functions, and $B$ is a symmetric bidifferential on $\Sigma^2$ with a double pole on the diagonal, vanishing residue, and normalized along $A$-cycles with respect to a chosen basis. Often the spectral curve is described instead by an affine equation
\begin{align}\label{spectralcurveP}
    P(x,y)=0,
\end{align}
where $P$ is a polynomial in two variables. In practice, one frequently allows more general spectral curves—for instance involving exponentials—corresponding to logarithmic poles of $x$ or $y$. 

In the classical setting of TR, one usually assumes $x$ and $y$ to have distinct ramification points. Then TR defines the multidifferentials as follows: one sets $\omega_{0,1}=y\,dx$ and $\omega_{0,2}=B$. For all $2g+n-1>0$, the higher differentials are given by
\begin{align}
  \label{eq:TR-intro}
  \omega_{g,n+1}(I,z)
  &= \sum_{p_i\in\operatorname{Ram}(x)}\Res\displaylimits_{q\to p_i} \left(\int^q_{p_i} B(\bullet,z)\right) \, 
  \tilde{\omega}_{g,n+1}(q,I)
\end{align}
where $\tilde{\omega}_{g,n+1}(p,I)$ is constructed from $\omega_{g',n'}$ with $2g'+n'-2<2g+n-2$.

Here $I=\{z_1,\dots,z_n\}$ denotes the set of coordinates on $\Sigma^n$. The set $\operatorname{Ram}(x)$ consists of the ramification points of $x$, that is all $p_i$ with $dx(p_i)=0$. The explicit form of $\tilde{\omega}_{g,n+1}(p,I)$ will not play any role in the rest of the article, we refer for details to \cite{Eynard:2007kz}.

%The free energies $F_g=\omega_{g,0}$ are not directly produced by \eqref{eq:TR-intro}. Instead, for $g>2$ one defines
%\begin{align}\label{freeenergydefinition}
%    F_g=\frac{1}{2-2g}\sum_{p_i\in\operatorname{Ram}(x)}
%  \Res\displaylimits_{q\to p_i}\omega_{g,1}(q)\,\Phi(q),
%\end{align}
%where $d\Phi(q)=\omega_{0,1}(q)$. The special cases $g=0,1$ require a more elaborate definition (see \cite{Eynard:2007kz}), which will not be needed here. More generally, TR admits a reduction formula
%\begin{align}
%    \omega_{g,n}(I)=\frac{1}{2-2g-n}\sum_{p_i\in\operatorname{Ram}(x)}
%  \Res\displaylimits_{q\to p_i}\omega_{g,n+1}(q,I)\,\Phi(q).
%\end{align}
%For instance, evaluating $F_2$ requires $\omega_{2,1}$, which depends on $\omega_{1,2}$ and $\omega_{1,1}$; computing $\omega_{1,2}$ in turn involves $\omega_{0,3}$. Hence, the algorithm rapidly becomes intricate for large $g$.

From the recursive definition several structural properties follow. For $2g+n-2>0$, the multidifferentials $\omega_{g,n}$:
\begin{itemize}
    \item are symmetric in their arguments,
    \item have poles only at ramification points of $x$,
    \item are residue-free, i.e. $\Res_{q\to p_i}\omega_{g,n}(q,I)=0$,
    \item satisfy homogeneity: $\omega_{0,1}\mapsto\lambda\omega_{0,1}$ implies $\omega_{g,n}\mapsto\lambda^{2-2g-n}\omega_{g,n}$,
    \item are invariant under
    \begin{align*}
        (x,y)&\mapsto (x,y+R(x)),\quad R\ \text{rational},\\
        (x,y)&\mapsto \Big(\tfrac{ax+b}{cx+d},\,y\tfrac{(cx+d)^2}{ad-bc}\Big).
    \end{align*}
\end{itemize}
The last property, the invariance transformation, reflects the underlying symplectic structure: the form $dx\wedge dy$ is preserved under the two transformations. Yet TR is not invariant under all symplectic transformations. For instance, interchanging $x$ and $y$ preserves $dx\wedge dy$ up to sign, but leads to a new family $\omega_{g,n}^\vee$, with poles now at the ramification points of $y$. Furthermore, one has $\omega_{0,1}^\vee=x\,dy$ and $\omega_{0,2}=B$.

An explicit formula between $\omega_{g,n}$ and $\omega_{g,n}^\vee$ for meromorphic $x,y$ was first conjectured in \cite{Borot:2021thu}, proven for $g=0$ in \cite{Hock:2022wer}, and established in full generality in \cite{Alexandrov:2022ydc}; equivalent formulations appear in \cite{Hock:2022pbw}. The general relation takes the form
\begin{align}\label{dualxyexpression}
    \omega_{g,n}=\text{Expr}_{g,n}\bigg(
        \big\{\omega_{h,m}^\vee\big\}_{2h+m-2\leq 2g+n-2},
        \{dy_i,dx_i\}_{i=1,\dots,n},
        \bigg\{\frac{dy_i\,dy_j}{(y_i-y_j)^2}\bigg\}_{i,j=1,\dots,n}
    \bigg),
\end{align}
where $\text{Expr}_{g,n}$ is a combinatorial algebraic expression involving just the dual differentials, the basic one-forms $dx_i,dy_i$, and a regularization term $\tfrac{dy_i\,dy_j}{(y_i-y_j)^2}$ that replaces $\omega_{0,2}^\vee$ on the diagonal. 
%For $n=1$, the first few terms of \eqref{dualxyexpression} are
%\begin{align}\label{dualxyexpressionfirstterms}
%    \omega_{g,1} = -\omega_{g,1}^\vee
%    + d\!\left(\frac{\omega_{g-1,2}^\vee+\tfrac{1}{2}\sum_{\substack{g_1+g_2=g\\ g_i>0}}\omega_{g_1,1}^\vee\,\omega_{g_2,1}^\vee}{dx\,dy}\right)
%    +\sum_{m=2}^{3g-1}\big(d\tfrac{1}{dx}\big)^m\Omega_{g,m},
%\end{align}
%where $\Omega_{g,m}$ are one-forms constructed from decorated graphs. Here $d\tfrac{1}{dx}$ denotes dividing a form by $dx$ followed by taking its exterior derivative, mapping one-forms to one-forms. 
We refer to \cite{Hock:2022pbw,Alexandrov:2022ydc} for the explicit expressions.  

The formula \eqref{dualxyexpression} is in fact universal: it holds for any two families of differentials $\omega_{g,n}$ and $\omega_{g,n}^\vee$ even if those are not generated by TR. This means that two families (with some additional conditions like no poles on the diagonal for $2g+n-2>0$) are related by the universal duality. The duality formula is indeed an involution. Building on this universal duality, a number of further developments around TR appeared already \cite{Alexandrov:2023jcj,Alexandrov:2023tgl,Alexandrov:2024ajj,Alexandrov:2024hgu,Alexandrov:2024tjo,Alexandrov:2024zku,Alexandrov:2025sap,Bouchard:2025rid,Hock:2023qii,Hock:2023dno,Hock:2025wlm,Hock:2025sxq,Banerjee:2025qgx}.

Most relevant for the present article is the notion of Log-TR, observed in \cite{Hock:2023dno} and refined in \cite{Alexandrov:2023tgl}. The explicit $x$–$y$ duality formula suggests that TR should be extended to account properly for logarithmic poles of $x,y$ (i.e. simple poles of $dx,dy$). Under suitable conditions—such as for generic framings of mirror curves of Calabi–Yau threefolds—Log-TR reduces to standard TR. This is precisely the setting of the remodeling conjecture \cite{Bouchard:2007ys}, later proved in \cite{Eynard:2012nj,Fang:2016svw}. For a fixed framing $f=0$, however, the remodeling conjecture fails \cite{Bouchard:2011ya}, but the problem is resolved in the Log-TR framework, which again enforces $x$–$y$ duality. Log-TR is also compatible with limits beyond those studied in \cite{Borot:2023wik}.

Let us briefly recall the definition. Denote by $a_1,\dots,a_M$ the logarithmic poles of $y$ that are not simultaneously poles of $dx$ (we will call those points \emph{log-vital} singular points). At each $a_i$, the differential $dy$ has non-vanishing residue $\tfrac{1}{t_i}\neq 0$. Then Log-TR is defined by $\omega_{0,1}=y\,dx$, $\omega_{0,2}=B$, and, for $2g+n-1>0$ we have for $\omega_{g,n+1}$,
\begin{align}
  \label{eq:TR-introLog}
  \omega_{g,n+1}(I,z)
  &=\sum_{p_i\in\operatorname{Ram}(x)}
  \Res\displaylimits_{q\to p_i}
  \left(\int^q_{p_i}\ B(\bullet,z)\right) \, 
  \tilde{\omega}_{g,n+1}(q,I)\\
  &\quad+\delta_{n,0}\sum_{i=1}^M
  \Res\displaylimits_{q\to a_i}
  \left(\int^q_{a_i}\ B(\bullet,z)\right)[\hbar^{2g}]
  \left(\frac{1}{t_i S(t_i\hbar\partial_{x(q)})}\log(q-a_i)\right)dx(q),\nonumber
\end{align}
where $S(t)=\tfrac{e^{t/2}-e^{-t/2}}{t}=\sum_{k=0}^{\infty} \frac{t^{2k}}{2^{2k}(2k+1)!}
= 1 + \frac{t^{2}}{24} +  \cdots$ understood as a formal expansion. Here,  $\tilde{\omega}_{g,n+1}(q,I)$ is the same as in \eqref{eq:TR-intro}, but will not be needed explicitly.

Throughout the rest of this article we will use the same notation $\omega_{g,n}$, with the understanding that they are defined via Log-TR. Compared with TR, only the $\omega_{g,1}$ sector is modified, but this modification propagates to all correlators through recursion. For $2g+n-2>0$, the resulting $\omega_{g,n}$ have poles only at ramification points of $x$, and for $n=1$ additionally at the log-vital singularities $a_i$. All $\omega_{g,n}$ remain symmetric and residue-free. Notably, even if $x$ is unramified, $\omega_{g,1}$ may still be nontrivial due to log-vital points of $y$.

The dual correlators $\omega_{g,n}^\vee$ are obtained by exchanging $x$ and $y$, i.e. taking residues at ramification points of $y$ and at the dual log-vital points of $x$. The families $\omega_{g,n}$ and $\omega_{g,n}^\vee$, both defined with Log-TR, are again related by the universal duality \eqref{dualxyexpression}. Some of the properties which are satisfied by TR in the setting of meromorphic $x,y$ are not necessarily satisfied in the more general setting of meromorphic $dx,dy$ and Log-TR. 

\medskip

An essential application of TR is the construction of a quantum spectral curve. The correlators $\omega_{g,n}$ (and analogously for Log-TR) determine a wave function $\psi_{x_0}(x)$ annihilated by a differential operator $\hat{P}_\hbar(\hat{x},\hat{y})$ in noncommuting variables. Here $\hat{x}=x\cdot$ and $\hat{y}=\hbar\,\tfrac{d}{dx}$ act on functions of $x$. The operator $\hat{P}_\hbar$ is a quantization of the original polynomial relation \eqref{spectralcurveP} in the semiclassical limit,
\begin{align*}
    \lim_{\hbar\to 0}\hat{P}_\hbar(x,y)=P(x,y).
\end{align*}

This idea traces back to the Airy case $P(x,y)=y^2-x$ \cite{Bergere:2009zm} but was already observed in matrix models before TR was even defined. For arbitrary genus-zero algebraic curves, a construction of the quantum spectral curve was derived in \cite{Bouchard:2016obz}. For higher-genus algebraic spectral curves, the construction of the wave function and quantum curve was developed in \cite{Iwaki:2019zeq,Marchal:2019bia,Eynard:2023fil,Eynard:2021sxg}. All these works, however, predate the introduction of Log-TR, and thus are not directly applicable to mirror curves of toric Calabi–Yau threefolds or to $A$-polynomials, which are not algebraic curves since they are in $\mathbb{C}\times \mathbb{C}$ rather than in $\mathbb{C}^\times \times \mathbb{C}^\times$ in logarithmic variables. Progress in this direction of $(\mathbb{C}^\times \times \mathbb{C}^\times)$-curves, including partial formulations of Log-TR, has been achieved in \cite{Marchal:2017ntz,Hock:2023dno,Hock:2025sxq}. In this article we build further upon those developments. Since we restrict ourselves to genus-zero spectral curves, the non-perturbative completion of the wave function (as introduced in \cite{Iwaki:2019zeq}) will not be required. Instead, we adopt the definition of \cite{Bouchard:2016obz}:
\begin{align}\label{wavefunction}
    \psi_{x_0}(x)=\exp\sum_{\substack{g\geq 0\\ n\geq 1}}
    \frac{\hbar^{2g+n-2}}{n!}
    \underbrace{\int_{x_0}^x\cdots\int_{x_0}^x}_{n\ \text{times}}
    \left(\omega_{g,n}-\frac{\delta_{g,0}\delta_{n,2}\,dx_1\,dx_2}{(x_1-x_2)^2}\right).
\end{align}
The wave function $\psi_{x_0}(x)$ depends also on the chosen base point $x_0$, and consequently the quantum curve $\hat{P}_\hbar$ does as well. Note that changing $x$ and $x_0$ is nothing else than changing the sign of $\hbar$ in the wave function, that is we have $\psi^{\hbar}_{x_0}(x)=\psi^{-\hbar}_x(x_0)$. Certain simplifications occur when $x_0$ is taken to be a singular point. Given a wave function constructed by the (Log-)TR multidifferentials $\omega_{g,n}$, the quantum spectral curve is the operator annihilating it (in general also depending on $x_0$)
\begin{align*}
    \hat{P}_\hbar(\hat{x},\hat{y};\hat{x}_0,\hat{y}_0)\psi_{x_0}(x)=0
\end{align*}
where the convention is used $\hat{y}_0=-\hbar \frac{d}{dx_0}$. With this, the quantum curve has the following symmetry 
$$\hat{P}_\hbar(\hat{x},\hat{y};\hat{x}_0,\hat{y}_0)=\hat{P}_{-\hbar}(\hat{x}_0,\hat{y}_0;\hat{x},\hat{y})$$
where we understand now $\hat{y}_0=\hbar\frac{d}{dx_0}$ and $\hat{y}=-\hbar\frac{d}{dx}$ in $\hat{P}_{-\hbar}$ since the wave function also changes the sign of $\hbar$ by going from $\psi_{x_0}(x)$ to $\psi_{x}(x_0)$.

\medskip

Let us mention the quite obvious realization of the $x$-$y$ duality for the quantum spectral curve and the wave function. What is the $x$-$y$ dual side of those two objects? Naively, one would guess that the dual wave function, defined by
\begin{align}\label{wavefunctiondual}
    \psi_{y_0}^\vee(y)=\exp\sum_{\substack{g\geq 0\\ n\geq 1}}
    \frac{\hbar^{2g+n-2}}{n!}
    \underbrace{\int_{y_0}^y\cdots\int_{y_0}^y}_{n\ \text{times}}
    \left(\omega^\vee_{g,n}-\frac{\delta_{g,0}\delta_{n,2}\,dy_1\,dy_2}{(y_1-y_2)^2}\right),
\end{align}
which is annihilated by the dual quantum spectral curve $\hat{P}^\vee_\hbar(\hat{y}^\vee,\hat{x}^\vee;\hat{y}^\vee_0,\hat{x}^\vee_0)$, is related to the $\psi_{x_0}(x)$ via Fourier transform. The reason for that is that the dual operators are $\hat{y}^\vee=\hbar \frac{d}{dy}$ and $\hat{x}^\vee=y\cdot $. Thus, the $x$-$y$ duality is on the level of wave functions nothing than the duality between momentum and position space in classical quantum mechanics \cite{Hock:2023dno}. However, the situation is not as easy as it seems, since $\psi_{x_0}(x)$ depends on two variables $x$ and the base point $x_0$. Therefore, a naive Fourier transform in $x$ is not always entirely correct. It was shown in  \cite{Weller:2024msm} that $\psi_{\infty}(x)$ and $\psi^\vee_{\infty}(y)$ are actually related by formal Fourier transform, that is $x_0$ and $y_0$ are simultaneously singular points of the spectral curve and the curve has genus zero. For general basepoints, the relation is given by \cite{Hock:2025sxq}
\begin{align}\label{ExtLaplace}
    \frac{\psi_{x(z_0)}(x(z))}{x(z)-x(z_0)} = \frac{i}{2\pi\hbar} \iint \frac{dy(\chi)dy(\chi_0)}{y(\chi)-y(\chi_0)} e^{\frac{x(z)y(\chi_0)-x(z_0)y(\chi)}{\hbar}} \psi^\vee_{y(\chi_0)}(y(\chi))
\end{align}
which is essentially a direct consequence of \cite{Alexandrov:2023jcj}. As mentioned before, for general base point, the quantum spectral curve will depend on $x_0$, thus it is of the form $\hat{P}_\hbar(\hat{x},\hat{y};\hat{x}_0,\hat{y}_0)$. Applying the transformation formula \eqref{ExtLaplace} to the quantum spectral curve, one finds that the two dual quantum spectral curves are related by 
\begin{align}\label{quatumcurvexydual1}
        & \hat{P}^\vee_\hbar(\hat{y}^\vee,\hat{x}^\vee;\hat{y}_0^\vee,\hat{x}_0^\vee)
        \\
        \notag 
        & = \hat{P}_{\hbar}\bigg(\hat{y}_0^\vee - \frac{\hbar}{\hat{x}^\vee - \hat{x}_0^\vee},
         \hat{x}_0^\vee - \frac{\hbar}{\hat{y}^\vee - \hat{y}_0^\vee};
        \hat{y}^\vee - \frac{\hbar}{\hat{x}^\vee - \hat{x}_0^\vee},
        \hat{x}^\vee - \frac{\hbar}{\hat{y}^\vee - \hat{y}_0^\vee}\bigg),
    \end{align}
where the right-hand side is literally understood as substitution (no normal ordering is applied). Note that the following convention is used $\hat{x}^\vee=y$, $\hat{y}^\vee=\hbar \frac{d}{dx^\vee}=\hbar\frac{d}{dy}$ and $\hat{x}_0^\vee=y_0$, $\hat{y}^\vee_0=-\hbar \frac{d}{dx^\vee_0}=-\hbar\frac{d}{dy_0}$. We can write equivalently
\begin{align}\label{quatumcurvexydual}
        &\hat{P}^\vee_{\hbar}\bigg(
         \hat{x}_0 - \frac{\hbar}{\hat{y} - \hat{y}_0},\hat{y}_0 - \frac{\hbar}{\hat{x} - \hat{x}_0};
        \hat{x} - \frac{\hbar}{\hat{y} - \hat{y}_0},\hat{y} - \frac{\hbar}{\hat{x} - \hat{x}_0}\bigg)\\
         =& \hat{P}_{\hbar}(\hat{x},\hat{y};\hat{x}_0,\hat{y}_0)\notag 
\end{align}

Most important for us is the fact that this duality transformation of the two quantum curves also holds in the framework of Log-TR. In other words, if two families $\omega_{g,n}$ and $\omega_{g,n}^\vee$ are related by the universal $x$-$y$ duality \eqref{dualxyexpression} (which is the case in Log-TR), then their perturbative wave functions are related by \eqref{ExtLaplace}. From this, one easily concludes that their quantum curves are related by \eqref{quatumcurvexydual1} or  \eqref{quatumcurvexydual}. We will apply this duality transformation for the quantum curve to the class of strip geometries of toric Calabi-Yau threefolds aka mirror curves.\\

The general idea of how we will proceed with the computations is the following. We consider a spectral curve whose dual side is trivial, so that all $\omega_{g,n}^\vee$ and the corresponding wave function are given explicitly. This yields the dual quantum curve $\hat{P}^\vee$, from which $\hat{P}$ can be obtained via \eqref{quatumcurvexydual}.

However, instead of deriving the wave function $\psi$ from $\psi^\vee$ through \eqref{ExtLaplace}, we will solve $\hat{P}\psi=0$ directly, obtaining a solution after resummation in a specific Stokes sector. This solution can then again be expressed as a Laplace transform as in \eqref{ExtLaplace}, but may correspond to another wave function $\psi^\vee$ resummed in a different sector.

In this way, we obtain an efficient procedure to derive the solution of the quantum curve. Nevertheless, the asymptotic $\hbar$-expansions of all such resummed solutions coincide. This reflects the general nature of resumming asymptotic series: we find a way to obtain the solution without resorting explicitly to resurgence theory.
\\

Finally, we want to mention that there is an explicit formula how to derive from the wave function $\psi_{x(z_0)}(x(z))$ with general basepoint all $\omega_{g,n}$ back. This formula is known under the name of \textit{determinantal formula}, first observed in \cite{Bergere:2009zm} for the Airy curve and further studied in \cite{Eynard:2023anx,Alexandrov:2023jcj}. Define the following kernel
\begin{align}\label{eq:bK-def}
 	K(z_1,z_2)& := \frac{\sqrt{dx_1\;dx_2}}{x_1-x_2}
 \exp \bigg(
 \sum\limits_{2g-2+n>0}\frac{\hbar^{2g-2+n}}{n!}
 		\int\limits_{z_2}^{z_1}\dots\int\limits_{z_2}^{z_1}\omega_{g,n}+
 \frac12
 		\int\limits_{z_2}^{z_1}\!\!\int\limits_{z_2}^{z_1}(\omega_{0,2}-\frac{dx_1dx_2}{(x_1-x_2)^2})\bigg),
\end{align}
where $x_i=x(z_i)$. The definition is related to the wave function by $K(z_1,z_2)=\psi_{x(z_2)}(x(z_1))\frac{\sqrt{dx_1\;dx_2}}{x_1-x_2}$. Then for $\omega_{n}=\sum_{g=0}^\infty\hbar^{2g+n-2}\omega_{g,n}$, the determinantal formula reads for a genus zero curve
\begin{align} \label{eq:omega-kappa-relation}
	\omega_1(z_1) & = \lim_{x_1'\to x_1} \Big(K(z_1,z_1')-\frac{\sqrt{dx_1dx'_1}}{x_1-x_1'}\Big); \\ \notag
	\omega_n(z_1,...,z_n)  & = (-1)^{n-1} \sum_{\sigma\in C_n} \prod_{i=1}^n
	K(z_{i},z_{\sigma(i)}), & n\geq 2,
\end{align}
where $C_n$ is the set of all permutations consisting of just one cycle of length $n$. In a sense, this formula takes the connected part of the determinant with matrix entries $(K(z_i,z_j))_{i,j}$.

\subsection{The dual Wave Function and the dual Quantum Spectral Curve for Strip Geometries}
The aim is to derive a quantum curve for the strip geometries used in \cite{Panfil:2018faz}, but from TR. We will do that by applying the recent developments in TR related to the $x$-$y$ duality and the Log-TR. The spectral curve reads
\begin{align}\label{spectralcurve}
    A(e^x,e^y)=(1-e^y)\prod_{j=1}^s(1-\beta_j e^y)+(-1)^f e^xe^{y(1+f)} \prod_{j=1}^r(1-\alpha_j e^y)=0
\end{align}
which has  for instance the parametrization
\begin{align}\label{parametricxy}
    x=&\log \bigg[(1-z)\frac{\prod_j (1-\beta_j z)}{\prod_j (1-\alpha_jz)}\bigg]-(f+1)\log z+f\log(-1)\\
    y=&\log z.
\end{align}
Replacing $e^x=X$ and $e^y=Y$, this curve becomes a curve in $\mathbb{C}^\times \times \mathbb{C}^\times$ since we have $\omega_{0,1}=\log Y\frac{dX}{X}$ which is the fundamental 1-form in $\mathbb{C}^\times \times \mathbb{C}^\times$. However, from TR perspective we prefer to work with $x,y$ rather than $X,Y$.

Let us analyze the $x$--$y$ dual side of TR. Due to the fact that $y$ is algebraically unramified (this means no algebraic ramification point, since the logarithmic ramification is not algebraic), we are in a very nice situation for the dual side. This means that all $\omega_{g,n}^\vee$ are trivial in the sense that they are given by an explicit expression. This explicit expression can be read off by the definition of Log-TR. 
To get $\omega_{g,n}^\vee$, we have to change in \eqref{eq:TR-introLog} the role of $x$ and $y$. The first two lines vanish and we are left for $2g+n-2>0$ with 
\begin{align}\nonumber
    \omega_{g,n}^\vee=& \delta_{n,1} \sum_{q\to a\in\{1,\frac{1}{\alpha_i},\frac{1}{\beta_j}\}}\Res\displaylimits_{q\to a}
  \int_{a}^q \omega_{0,2}(z,\bullet)[\hbar^{2g}]
  \left(\frac{1}{t_a S(t_a \hbar\partial_{y(q)})}\log(q-a)\right)dy(q)\\\label{omgdual}
    =&\delta_{n,1}[\hbar^{2g}]dy \frac{1}{S(\hbar \partial_y)}x,
\end{align}
where $t_a=\pm 1$, $S(t)=\frac{e^{\frac{t}{2}}-e^{-\frac{t}{2}}}{t}$ and $x,y$ are given by \eqref{parametricxy}. The sign of $t_a$ depends on the residue of $dx$ at $a\in\{1,\frac{1}{\alpha_i},\frac{1}{\beta_j}\}$ which is $+1$ for $a\in\{1,\frac{1}{\beta_j}\}$ and $-1$ for $a\in\{\frac{1}{\alpha_i}\}$. Going from the first to the second line is a straightforward computation by observing that the second line has only poles located at $\{1,\frac{1}{\alpha_i},\frac{1}{\beta_j}\}$, no residues and vanishes at infinity. Together with
\begin{align*}
    \omega_{0,2}^\vee=\frac{dz_1\,dz_2}{(z_1-z_2)^2},
\end{align*}
all dual $\omega_{g,n}^\vee$ are known explicitly. 

At this point, we want to emphasize that the differentials provided by TR are in a sense coefficients of a perturbative (asymptotic) series in $\hbar$ \cite{Borot:2024lof}. Here, we really think of a series with radius of convergence zero. 
If we want to write down a wave function roughly speaking as $\psi^\vee=\exp \sum \int \omega_{g,n}^\vee$ as defined in \eqref{wavefunctiondual}, there are ambiguities in resumming the divergent series over $\hbar$. Being more explicitly, all integrated $\omega_{g,n}^\vee$ with base point $y_0=y(z_0)$ with $z_0=\infty$ are (after proper regularization at $z_0$ for $\omega_{0,1}^\vee$ and $\omega_{0,2}^\vee$)
\begin{align}\nonumber
    &\int^z_{z_0=\infty}\int^z_{z_0=\infty} \omega_{0,2}^\vee-\frac{dy_1\,dy_2}{(y_1-y_2)^2}\bigg\vert_{reg}=\log\left(\frac{z_1-z_2}{y(z_1)-y(z_2)}\right)\bigg\vert^{z_1,z_2\to z}_{z_1,z_2\to z_0,reg}\\
    &=\log\left(\frac{1}{y'(z)}\right)=\log(z),\\\nonumber
    &\sum_{g=0}^\infty\hbar^{2g-1}\int^z_{z_0=\infty}\omega_{g,1}^\vee=e^{\frac{\hbar}{2}\partial_y}\int \bigg(\frac{\partial_y}{e^{\hbar\partial_y}-1}x\bigg)dy=e^{\frac{\hbar}{2}\partial_y}\frac{1}{\hbar}\sum_{n=0}^\infty\hbar^n\frac{B_n}{n!}\partial_y^{n-1}x\\\label{omg1veepert}
    &=-\frac{1+f}{2 \hbar} \log(z)^2 -e^{\frac{\hbar}{2}\partial_y}\frac{1}{\hbar}\sum_{n=0}^\infty\hbar^n\frac{B_n}{n!}\left(\mathrm{Li}_{2-n}(z)+\sum_j \mathrm{Li}_{2-n}(\beta_j z)-\sum_j \mathrm{Li}_{2-n}(\alpha_j z)   \right)\\\nonumber
    &+ \frac{1}{\hbar}f\log(-1)\log z
\end{align}
understood as a formal expansion in $\hbar$. We have used identities of the polylogarithm of the form $\partial_y\mathrm{Li}_n(a z)=\mathrm{Li}_{n-1}(a z)$. 

Resumming divergent series can be performed via resurgence, where first the Borel transform has to be derived and then, roughly speaking, a Laplace transform of the Borel transform leads to a possible analytic function in $\hbar$ away from zero. The function itself depends on integration contour or more precisely the integration direction for which the Laplace transform is computed. For the $\omega_{g,n}^\vee$, which occur in our setting, the resulting wave function has the same asymptotic behavior as the quantum dilogarithm \cite{Faddeev:1993rs}. The asymptotics resummation via resurgence of the quantum dilogarithm was studied in \cite{MR1982715,MR4394512}. In the following, we will use one possible Borel resumed function which does not coincide with the quantum dilogarithm but correspond to the solution in another sector. Going from one sector to another, wall-crossing-like formulas can be applied, also known as Stokes phenomenon. The sector we are interested in corresponds to the integration ray with angle $\theta=\frac{\pi}{2}$ in \cite{MR4394512}. Thus, we can rewrite the solution in this sector
\begin{align*}
    \sum_{g=0}^\infty\hbar^{2g-1}\int^z_{z_0=\infty}\omega_{g,1}^\vee=&-\frac{1+f}{2\hbar} \log(z)^2-\log (zq^{1/2},q)_\infty-\sum_j\log (z\beta_j q^{1/2},q)_\infty\\
    &+\sum_j\log (z\alpha_j q^{1/2},q)_\infty
    +\frac{1}{\hbar} f\log(-1)\log z,
\end{align*}
where $q=e^\hbar$ and $(z,q)_\infty=\prod_{i=0}^\infty(1-zq^{i})$ an infinite $q$-Pochhammer symbol. We can conclude that the wave function $\psi_{\infty}^\vee(y(z))$ takes in one of the sectors the form
\begin{align}\nonumber
    \psi_{\infty}^\vee(y(z))=&\exp  \sum_{g=0}^\infty\hbar^{2g-1}\int^z_{z_0}\omega_{g,1}^\vee+\frac{1}{2}\int^z_{z_0}\int^z_{z_0} \omega_{0,2}^\vee-\frac{dy_1\,dy_2}{(y_1-y_2)^2}\bigg\vert_{reg}\\\label{dualwavepoch}
    =&\sqrt{z} \frac{\prod_j (z\alpha_j q^{1/2},q)_\infty}{(zq^{1/2},q)_\infty \prod_j (z\beta_j q^{1/2},q)_\infty}e^{-\frac{1+f}{2\hbar }\log(z)^2 } (-z)^{\frac{f}{\hbar}},
\end{align}
with $q=e^\hbar$.
\begin{remark}
    As explained above, it is also possible to integrate the Laplace transform of the Borel resummation along another direction. A second natural direction is $\theta=0$, see \cite{MR4394512}. The resulting wave function would be expressed in terms of the Quantum Dilogarithm
    \begin{align}\label{dilogpsivee}
        \psi^{\vee,\theta=0}_{\infty}(y)=e^{y/2}e^{-\frac{1+f}{2\hbar }y^2 } (-1)^{\frac{fy}{\hbar}}\frac{\prod_j \Phi_b(\frac{y+\log (-\alpha_j)}{2\pi b})}{\Phi_b(\frac{y+\log(-1)}{2\pi b})\prod_j \Phi_b(\frac{y+\log(- \beta_j)}{2\pi b})},
    \end{align}
    with $b^2=\frac{\hbar}{2\pi i}$ and $\Phi_b(y)$ the quantum dilogarithm of Faddeev. This can be written as a ratio of two infinite $q$-Pochhammer symbols. 

    Remarkably, for any integration direction $\theta$ in the Borel plane, the resulting wave function $\psi^{\vee,\theta}_{\infty}(y)$ can be written as ratio or products of infinite Pochhammer symbols. Thus, the $x$--$y$ dual side of the strip geometries admits a very simple representation. 
\end{remark}
Let us now derive the $q$-difference equation satisfied by the dual wave function $\psi^\vee_\infty(y)$. First, insert the definition of $\omega_{g,1}^\vee$ in the definition of the wave function. Instead of expanding it as in \eqref{omg1veepert}, we want to invert the difference operator, and get
\begin{align*}
    (e^{\frac{\hbar}{2} \partial_y}-e^{-\frac{\hbar}{2} \partial_y})\log \psi_\infty^\vee(y)=(e^{\frac{\hbar}{2} \partial_y}-e^{-\frac{\hbar}{2} \partial_y})\bigg(\frac{y}{2}\bigg) +x(y),
\end{align*}
where the first term on the RHS comes from $\omega^\vee_{0,2}$ and $x(y)$ literally means the solving the spectral curve wrt $x$. Let the difference operator act and take the exponential, we derive
\begin{align}\label{differencey}
    \psi_\infty^\vee\left(y+\frac{\hbar}{2}\right)= e^{\frac{\hbar}{2}}\frac{(1-e^y)\prod_{j=1}^s(1-\beta_j e^y)}{(-1)^{f+1} e^{y(1+f)} \prod_{j=1}^r(1-\alpha_j e^y)}\psi_\infty^\vee\left(y-\frac{\hbar}{2}\right).
\end{align}
It is easy to verify that \eqref{dualwavepoch} and \eqref{dilogpsivee} are solutions of this $q$-difference equation. Note that $q$-difference equations are usually written in the variable $Y=e^y$, where $e^{y+\frac{\hbar}{2}}=Yq^{1/2}$.

\begin{remark}\label{rem:factordual}
    We want to emphasize that the base point (which was chosen to be $z_0=\infty$) affects the quantum curve in a simple way. The reason for this lies in the fact that $\omega_{g,n}^\vee$ with $n>1$ vanish. This means that the wave function for an arbitrary base point $z_0$ factorizes in three factors depending on $z$, $z_0$, and one coming from $\omega_{0,2}^\vee$, since the integration $\int_{z_0}^z\omega_{g,1}^\vee$ gives just the difference of two terms depending on $z$ and $z_0$, respectively. For $\int_{z_0}^z...\int_{z_0}^z\omega_{g,n>1}$, the terms mix and depend on $z$ and $z_0$. However, one mixing appears from $\omega_{0,2}^\vee$. Thus, the dual wave function for arbitrary base point factorizes as follows:
    \begin{align}
        \psi_{y_0}^\vee(y)=\frac{y(z)-y(z_0)}{e^{y(z)}-e^{y(z_0)}}\psi^\vee(y(z))\psi^\vee(y(z_0)).
    \end{align}
    The prefactor comes from the integration of $\omega_{0,2}$. Here, $\psi^\vee(y(z))$ satisfies the same $q$-difference equation \eqref{differencey}. The factor $\psi^\vee(y(z_0))$ can be neglected since we are looking for $q$-difference equation in $y(z)$. The impact of the prefactor is straightforward to determine and importantly independent of $\hbar$.

    Nevertheless, when going to the dual side and deriving the wave function $\psi(x)$ and its $q$-difference equation or equivalently the quantum curve, this choice of the base point becomes important. This can be seen from \eqref{quatumcurvexydual} which explains how the quantum curves transforms by going to its dual side, which now depends on the basepoint!
\end{remark}
With $\hat{x}^\vee=y\cdot$ and $\hat{y}^\vee=\hbar \frac{d}{dy}$, the $q$-difference equation \eqref{differencey} reads as the dual quantum curve with base point at $z_0=\infty$
\begin{align}\label{dualquantumcurve}
    \hat{A}^\vee_\hbar(\hat{y}^\vee,\hat{x}^\vee)=e^{\frac{\hat{y}^\vee}{2}}-e^{\frac{\hbar}{2}}\frac{(1-e^{\hat{x}^\vee})\prod_{j=1}^s(1-\beta_j e^{\hat{x}^\vee})}{(-1)^{f+1} e^{{\hat{x}^\vee}(1+f)} \prod_{j=1}^r(1-\alpha_j e^{\hat{x}^\vee})}e^{-\frac{\hat{y}^\vee}{2}}.
\end{align}
\begin{remark}\label{rem:factordual2}
    For general base point $\psi^\vee_{y_0}(y)$ as in Remark \ref{rem:factordual}, the $q$-difference equation reads\footnote{As mentioned in Remark \ref{remarklinearequation}, for all strip geometries the dual wave function satisfies for any basepoint a linear $q$-difference equation, which we have just proved.} 
    \begin{align}\label{differenceyy0}
    \psi_{y_0}^\vee\left(y+\frac{\hbar}{2}\right)= e^{\frac{\hbar}{2}}\frac{y+\hbar/2-y_0}{y-\hbar/2-y_0}\frac{e^{y-\hbar/2}-e^{y_0}}{e^{y+\hbar/2}-e^{y_0}}\frac{(1-e^y)\prod_{j=1}^s(1-\beta_j e^y)}{(-1)^{f+1} e^{y(1+f)} \prod_{j=1}^r(1-\alpha_j e^y)}\psi_{y_0}^\vee\left(y-\frac{\hbar}{2}\right).
\end{align}
We can extract the dual quantum curve and observe the explicit dependence on $y_0$ which is written as operator $\hat{x}^\vee_0$:
\begin{align}\label{subz010A}
    &\hat{A}^\vee_\hbar(\hat{y}^\vee,\hat{x}^\vee;\hat{y}^\vee_0,\hat{x}^\vee_0)\\\nonumber
    =&e^{\frac{\hat{y}^\vee}{2}}-e^{\hbar/2}\frac{\hat{x}^\vee+\hbar/2-\hat{x}^\vee_0}{\hat{x}^\vee-\hbar/2-\hat{x}^\vee_0}\frac{e^{\hat{x}^\vee-\hbar/2}-e^{\hat{x}^\vee_0}}{e^{\hat{x}^\vee+\hbar/2}-e^{\hat{x}^\vee_0}}\frac{(1-e^{\hat{x}^\vee})\prod_{j=1}^s(1-\beta_j e^{\hat{x}^\vee})}{(-1)^{f+1} e^{{\hat{x}^\vee}(1+f)} \prod_{j=1}^r(1-\alpha_j e^{\hat{x}^\vee})}e^{-\frac{\hat{y}^\vee}{2}}.
\end{align}
\end{remark}

\subsection{The Wave Function and the Quantum Spectral Curve for Strip Geometry} \label{sec.wavestrip}
The last step consists of going from the dual side (from the dual wave function $\psi^\vee$ and the dual quantum curve $\hat{A}^\vee_\hbar$) to the wave function $\psi$ and the quantum curve $\hat{A}_\hbar$. Here, the choice of the base point $z_0$ becomes important. This can be seen from two perspectives: Firstly, all $\omega_{g,n}$ are nontrivial, thus their integrals appearing in the definition of the wave function have mixing terms which is clearly different for the dual side, see Remark \ref{rem:factordual}. Secondly, going from the dual quantum curve $\hat{A}^\vee$ to $\hat{A}$ via \eqref{quatumcurvexydual}, we observe that even if $\hat{A}^\vee$ would be independent of $\hat{x}^\vee_0$ and $\hat{y}^\vee_0$, the arguments have nontrivial appearance of $\hat{x}_0$ and $\hat{y}_0$.

\subsubsection{Quantum Spectral Curve for Strip Geometry with base point $z_0=\infty$}
Going to the dual quantum curve, it is much simpler if the base point $z_0$ is chosen to be simultaneously a singular point for $x$ and $y$. This is the reason for us choosing $z_0=\infty$, since it is a singular point for $x$ and $y$. In this situation, the duality in going from $\hat{A}^\vee$ to $\hat{A}$ is literally a replacing of $\hat{y}^\vee$ by $\hat{x}$, $\hat{x}^\vee$ by $\hat{y}$ and changing the sign of $\hbar$. The change of the $\hbar$-sign comes from the fact that in \eqref{quatumcurvexydual} additionally the first two arguments are interchanged with the third and fourth. For $z_0$ being a singular point of $x$ and $y$, the duality transformation between $\hat{A}^\vee_\hbar$ and $\hat{A}_\hbar$ was already derived in \cite{Weller:2024msm}. The generalization to arbitrary base point is \eqref{quatumcurvexydual}.

Applying this transformation to \eqref{dualquantumcurve}, we find
\begin{align}
    \hat{A}_\hbar(\hat{x},\hat{y})=e^{\frac{\hat{x}}{2}}-e^{-\frac{\hbar}{2}}\frac{(1-e^{\hat{y}})\prod_{j=1}^s(1-\beta_j e^{\hat{y}})}{(-1)^{f+1} e^{{\hat{y}}(1+f)} \prod_{j=1}^r(1-\alpha_j e^{\hat{y}})}e^{-\frac{\hat{x}}{2}}.
\end{align}
Quantum curves can be multiplied from the left by further operators which does not change the $q$-difference equation. In this way, we will cancel the denominator. Furthermore, multiplying from the left by $e^{\frac{\hat{x}}{2}}$ and commuting the $\hat{x}$ and $\hat{y}$ operators, we find
\begin{align}\label{eq:TRAhatbpinfty}
    \hat{A}_\hbar(\hat{x},\hat{y})=e^{\hat{x}}(-1)^{f+1} q^{1+\frac{f}{2}} e^{{\hat{y}}(1+f)} \prod_{j=1}^r(1-\alpha_j q^{\frac{1}{2}}e^{\hat{y}})-(1-q^{-\frac{1}{2}}e^{\hat{y}})\prod_{j=1}^s(1-\beta_j q^{-\frac{1}{2}}e^{\hat{y}})
\end{align}
with $q=e^\hbar$.
Writing the solution for the $q$-difference equation as $\psi^{(\infty)}_f(x) = \sum_{n\geq 0} c_n e^{nx}$, {\it for small $e^x$} \footnote{For certain choices of $f$ and complex structure parameters of the curve, this might not be true when $y=\infty$. For those cases, we have to do the analysis separately. This was discussed in \textsection\ref{sec:intro}} which is the valid regime in this case, one gets the following recursion
\begin{equation}
c_n (1-q^{n-\frac12})\prod _{j=1}^s (1-\beta_j q^{n-\frac12}) = c_{n-1} (-1)^{f+1} q^{1+f/2} q^{(n-1)(1+f)} \prod_{j=1}^r (1-\alpha_j q^{n-\frac12})
\end{equation}
for $n\geq 1$. We also choose here a scheme for resummation so as to obtain a convergent expression in $q$, implicitly \cite{MR4394512,Grassi:2022zuk,Alim:2022oll}. \footnote{We ensure that $|q|<1$ and this implies $\arg (\hbar)\in [\pi/2,3\pi/2]$. Our choice of $q=e^\hbar$ is such that $\arg (\hbar)$ lies in one of the relevant Stokes sector. }Then Setting $c_0 = 1$, the solution is then 
\begin{equation}
\label{eq:GWfnbpinfty}
\psi^{(\infty)}_f(x) = \sum_{n\geq 0} (-1)^{n(f+1)} q^{\frac{n^2}{2}(1+f) + \frac{n}{2}}
\frac{\prod_{j=1}^r (q^{\frac12}\alpha_j;q)_n}{(q^\frac12;q)_n\prod_{j=1}^s (q^\frac12\beta_j;q)_n} \,e^{nx}
\end{equation}

\subsubsection{Quantum Spectral Curve for Strip Geometry with base point $z_0=1$}
For completeness, we want to add one further example of a quantum curve with a nontrivial base point. This means that a base point is chosen which is not simultaneously a singular point for $x$ and $y$. A good choice would be $z_0$ which is still singular for $x$, but $y$ vanishes, $y(z_0=1)=0$. Following Remark \ref{rem:factordual} and \ref{rem:factordual2}, the dual wave function $\psi^\vee_{y(z_0=1)}(y)$ with base point $z_0=1$ is annihilated by the quantum curve 
\begin{align}\label{subz010}
    e^{\frac{\hat{y}^\vee}{2}}-e^{\hbar/2}\frac{\hat{x}^\vee+\hbar/2}{\hat{x}^\vee-\hbar/2}\frac{e^{\hat{x}^\vee-\hbar/2}-1}{e^{\hat{x}^\vee+\hbar/2}-1}\frac{(1-e^{\hat{x}^\vee})\prod_{j=1}^s(1-\beta_j e^{\hat{x}^\vee})}{(-1)^{f+1} e^{{\hat{x}^\vee}(1+f)} \prod_{j=1}^r(1-\alpha_j e^{\hat{x}^\vee})}e^{-\frac{\hat{y}^\vee}{2}}.
\end{align}
The difference of this dual quantum curve entirely comes from the integration of $\omega^\vee_{0,2}$ with the integration divisor $z_0=1$. 

In the next step, we have to go to the dual side by using \eqref{quatumcurvexydual} which relates the two dual quantum curves. Importantly, here since $z_0$ is not a singular point of $y$, we cannot just replace $\hat{y}^\vee$ by $\hat{x}$. However since $z_0$ is still singular for $x$, we can still replace (note the $\hbar$-sign change)
\begin{align}\label{subz011}
    \hat{x}^\vee \to \hat{y}+\frac{\hbar }{\hat{x}-\hat{x}_0}= \hat{y}.
\end{align}
For the replacement of $\hat{y}^\vee$, we find
\begin{align*}
\hat{y}^\vee\to \hat{x}+\frac{\hbar}{\hat{y}-\hat{y}_0}.
\end{align*}
Here, $\hat{y}_0$ which is the derivative wrt $x_0=x(z=1)=\infty$ should be treated as explained in \cite[Lem. 3.8]{Hock:2025sxq}. It can be replaced by $y_0=y(z=1)=0.$ Thus, we substitute 
\begin{align}\label{subz012}
    \hat{y}^\vee\to \hat{x}+\frac{\hbar }{\hat{y}}.
\end{align}
As mentioned, the substitution is literally a substitute, no normal ordering is done before. Thus, we arrive at an operator that has $\hat{y}$ in the denominator of an exponential which is quite unnatural, but amazing simplifications appear of the form
\begin{align}\label{subz013}
    e^{\pm \frac{1}{2}\big(\hat{x}+\frac{\hbar }{\hat{y}}\big)}=e^{\pm \frac{1}{2}\big(\hat{x}\frac{\hat{y}}{\hat{y}}+\frac{\hbar }{\hat{y}}\big)}=e^{\pm \frac{1}{2}\big(\hat{y}\hat{x}\frac{1}{\hat{y}}\big)}=\hat{y}e^{\pm \frac{1}{2}\hat{x}}\frac{1}{\hat{y}}=\frac{\hat{y}}{\hat{y}\pm \hbar/2}e^{\pm \frac{1}{2}\hat{x}}.
\end{align}
Finally, inserting \eqref{subz011}, \eqref{subz012} and  \eqref{subz013} in \eqref{subz010}, we derive via \eqref{quatumcurvexydual} (and after multiplying from the right with operators to cancel the denominators)
\begin{align}\notag
    \hat{A}_{\hbar}(\hat{x},\hat{y})=&e^{\frac{\hat{x}}{2}}(e^{\hat{y}}q^{-\frac{1}{2}}-1)q^{\frac{1}{2}}(-1)^{f+1} e^{{\hat{y}}(1+f)} \prod_{j=1}^r(1-\alpha_j e^{\hat{y}})e^{\frac{\hat{x}}{2}}\\ \label{eq:Ahatbp1}
    &-e^{\frac{\hat{x}}{2}}(e^{\hat{y}}q^{\frac{1}{2}}-1)(1-e^{\hat{y}})\prod_{j=1}^s(1-\beta_j e^{\hat{y}})e^{-\frac{\hat{x}}{2}}\notag\\
    =&e^{\hat{x}}(e^{\hat{y}}-1)(-1)^{f+1} q^{1+\frac{f}{2}}e^{{\hat{y}}(1+f)} \prod_{j=1}^r(1-\alpha_j q^{\frac{1}{2}}e^{\hat{y}})\\
    &-(e^{\hat{y}}-1)(1-q^{-\frac{1}{2}}e^{\hat{y}})\prod_{j=1}^s(1-\beta_jq^{-\frac{1}{2}} e^{\hat{y}})\notag
\end{align}
In the semi-classical limit $q\to 1$, this reduces to the spectral curve multiplied with $(e^y-1)$, but the quantum curve itself does not factorize. Note that this factor corresponds to the base point $(e^y-1)=0$ which becomes quantized  nontrivially in this case. 

\begin{remark}
    In the quantization of $A$-polynomials, which appears in the context of the AJ-conjecture \cite{MR2172488}, the semiclassical limit has usually several connected components consisting of abelian and non-abelian parts. The non-abelian part is the nontrivial part of $A$-polynomial. From the example above, one might ask that the abelian components will correspond to the base point which is chosen to construct the quantum curve from TR annihilating the colored Jones polynomial (wave function) \cite{Borot:2012cw}.
\end{remark}

As mentioned, unlike the previous case, here $y = \log z$ remains finite. Thus the solution holds for a different point on the mirror curve. For this case we also choose our resummation scheme consistent with the case before, namely the same $\arg (\hbar)$. Plugging an ansatz $\psi^{(1)}_f(x) = \sum_{n\geq 0} c_n e^{nx}$ once again (assuming $e^x$ to be small and necessarily $f\geq -1$), one derives the recursion 
\begin{equation}
\begin{split}
& c_n (1-q^n) (1-q^{n-\frac12})\prod _{j=1}^s (1-\beta_j q^{n-\frac12}) 
\\ & 
= c_{n-1} (-1)^{f+1} q^{1+f/2} q^{(n-1)(1+f)}(1-q^{n-1}) \prod_{j=1}^r (1-\alpha_j q^{n-\frac12})
\end{split}
\end{equation}
for $n\geq 1$. Then Setting $c_0 = 1$, the solution is then 
\begin{equation}
\label{eq:GWfnbp1}
\psi^{(1)}_f(x) = \sum_{n\geq 0} (-1)^{n(f+1)} q^{\frac{n^2}{2}(1+f) + \frac{n}{2}}
\frac{(q;q)_n\prod_{j=1}^r (q^{\frac12}\alpha_j;q)_n}{(q^\frac12;q)_n (q^2;q)_n\prod_{j=1}^s (q^\frac12\beta_j;q)_n} \,e^{nx}
\end{equation}
In the summand the appearance of an extra ratio of Pochhammers signify the wall-crossing of the BPS states compared to \eqref{eq:GWfnbpinfty}.

\section{Relations to open DT invariants and quiver generating series}

\label{sec:openDT}
This section aims to connect the quantum curve derived from TR for strip geometries in section \textsection \ref{sec:TRStrip} to other approaches studied in theoretical physics and mathematical physics. The nature of those relations comes fundamentally from the spectral curves being a curve on $\mathbb{C}^\times\times \mathbb{C}^\times$. Those can be for instance mirror curves or more curves appearing in more general contexts. This puts the following conclusions in much more wider context, than what has been studied in this article. 

We analyze the results of Section \ref{sec.wavestrip} from three angles as follows. 
\begin{itemize}
    \item Symmetric quiver : They naturally appear in the context of studying the open DT invariants \eqref{eq:DTfnintro}. In the work  of Efimov \cite{Efimov_2012}, expanding on the works of Kontsevich and Soibelman \cite{kontsevich2011cohomologicalhallalgebraexponential}, it was shown that to a finite symmetric quiver, one can associate a free cohomological Hall algebra. From the associated generating function, the quantum DT invariants were defined through \eqref{eq:DTfnintro}. The equivalence of this equation with Efimov's definition was indicated in \cite{Panfil:2018faz}, for the case of strip geometries. The DT invariants are labeled by the complex structure parameters of the mirror curve, the choice of a point on the mirror curve $x$ (which equivalently is where the Aganagic-Vafa brane sits on the toric diagram on the A-model side) and another integer parameter $j\in \mathbb{Z}$. The stability data of the representation is encoded by $x$ and the complex structure parameters, provided we keep the $\vartheta= \arg (\hbar)$ fixed. Then the quiver is equivalently defined in terms of the generalized Nahm sum expressions, which are the main objects of study in subsection \ref{sec:quivers}. 

    We briefly recall, how we arrive there. We start with Log-TR on the curve \eqref{spectralcurveintro}. With the aid of $x$--$y$ duality, we obtain a formal $\hbar$ expansion of the wave function which depends on the choice of the basepoint. Instead of resumming this expression, we take an alternative route. Through Log-TR one can also define ``quantum curve'', the form of which also depends on the choices of the basepoints. Then one can directly solve the difference equation around special points on the $x$-plane, which take the form of the generalized Nahm sum, defining the open DT invariants. We contrast how the Nahm sums depend on the choice of basepoints \eqref{eq:GWfnbpinfty} and \eqref{eq:GWfnbp1}. Clearly, the corresponding DT invariants also change. 
    
    \item Next comes the relations to the exponential networks which we briefly recall in subsection \ref{sec:expnet} following \cite{Eager:2016yxd, Banerjee:2018syt,Banerjee:2019apt,Banerjee:2020moh,Banerjee:2024smk}. The connection of exponential networks to these generalized Nahm sums were also considered in \cite{Grassi:2022zuk,Alim:2022oll}. In fact with exponential networks one studies two types of wall-crossing of BPS states, one of them is for fixed $\vartheta$ and varying $x$, where the network defining equation is given by \eqref{netweq}. In the second step one varies $\vartheta$. In our setup for this paper we are interested only with varying $x$ and see how the  spectrum of BPS states changes. Using exponential networks, starting from the difference equations, through the package of resurgence, one can construct the wave functions of this paper, which was in fact done in \cite{Grassi:2022zuk, Alim:2022oll} for $\mathbb{C}^3$ and resolved conifold, for $f=-1$. 

    Discussions in subsection \ref{sec:quivers} and \ref{sec:expnet} illustrate the main point of this work, how open GW invariants are related to open DT invariants. \footnote{Studying for higher genus mirror curves to understand more details will be certainly the next step towards having a better understanding of this correspondence.} However, we would like to propose that first step of the {\it nonabelianization procedure} defined in \cite{Banerjee:2018syt},  allows for computing a WKB solution for the associated difference equation (see also \cite{DelMonte:2024dcr}, as function of $x$, but asymptotic in $\hbar$. This result is expected to match with our asymptotic expansion of the wave function in $\hbar$. This fact could be tied up as connection between TR and exponential networks. 
    
    \item Then in subsection \ref{sec:qBarnes}, for these generalized Nahm sums, we give the $q$-Barnes type integral representations. Such integrals can be written in the form $\int_\Gamma dZ e^{\hbar^{-1} W(X,Z;\hbar)} g(Z;\hbar)$. \footnote{The $T^{[3D]}(L)$ theory as was said admits a description of a 3D $\mathcal{N=2}$ $U(1)$ Landau-Ginzburg model. This $W(X,Z;\hbar)$ can be thought of as the Landau-Ginzburg potential.} These functions $W(X,Z;\hbar)$ define the so called massive $tt^*$ theories with isolated critical points. The 2D versions of these theories have been under attention since the works of Cecotti-Vafa \cite{Cecotti:1992rm}, where one could also understand the wall-crossing of the 2D BPS states in terms of wall-crossing of Lefschetz-thimbles through Picard Lefschetz theory. For the present context, for 3D $tt^*$ systems, some aspects were studied in \cite{Banerjee:2018syt}. Study of similar wall-crossing appeared also naturally in \cite{kontsevich2024holomorphicfloertheoryi}. For this present class of examples of strip geometries, the $q$-Barnes integrals package all of these aspects together. We make one more observation: with proper choice of contour, the $q$-Barnes integrals are of the form of partition functions of 3D Gauged Linear Sigma Models (GLSM) on $S^1\times_q D^2$, where $D^2$ is a disk. In fact reduction to 2D has close connection to the hemisphere partition function of \cite{Hori:2013ika}, corresponding to 2D $\mathcal{N}=(2,2)$ GLSM on a disk. Furthermore, the hemisphere partition functions satisfy certain quantum difference equations which seemingly have nature as the difference equations studied here. In \cite{Banerjee:wwp}, at least in certain cases we will identify the generalized Nahm sums appearing in this paper and the $q$-Barnes representation as 3D GLSM partition functions \cite{Yoshida:2014ssa}. Some aspects to the connections to difference equations that we studied here were already hinted at \cite{Dimofte:2017tpi}. Studying the larger class presented in this paper however requires further investigations. 

    One more connection becomes apparent \cite{Panfil:2018faz, Banerjee:2020dqq}, if we use the fact $\log (X;q)_d \sim \frac{1}{\hbar} (\mathrm{Li}_2(X) - \mathrm{Li}_2(X q^d) )$, to rewrite \eqref{eq:DTfnintro} as (converting the sum to integral and changing $Z_i = e^{\hbar d_i}$)
    \begin{equation}
    P_C(X_1,...,X_n) \sim \int \frac{dZ_1 ...dZ_m}{Z_1...Z_m} \exp\bigg[\frac{1}{\hbar} \widetilde{W}(X,Z)\bigg],
    \end{equation}
where once again extremizing with respect to each $Z_i$, as $e^{Z_i\frac{\partial \widetilde{W}}{\partial Z_i}} = 1$ we recover the curve (in our context $n=1$). Now writing the generalized Nahm sum in the form of $q$-Barnes function, we also have an exact expression for the above integral, in terms of integral over one variable, which directly related to $T^{[3D]}(L)$, alluded to above. 

Finally, the integrands of the $q$-Barnes integrals satisfy a linear difference equation which is much easier to work with. Knowing that solution, one could compute the wavefunction for any $f$ by Fourier-Laplace transform. Changing the framing $f$, the quiver matrix changes \cite{Panfil:2018faz}. This change is more tractable in terms of these $q$-Barnes integrals.

\end{itemize}

We recall the parametrization of our curve as \eqref{parametricxy}
\begin{align}\label{parametricxy:intro}
    x=&\log \bigg[(1-z)\frac{\prod_j (1-\beta_j z)}{\prod_j (1-\alpha_jz)}\bigg]-(f+1)\log z+{f\log(-1)}\\
    y=&\log z.
\end{align}
\begin{itemize}

\item Choosing the basepoint $z_0=\infty$, working in the corresponding neighborhood, one derives the quantum curve \eqref{eq:TRAhatbpinfty}
\begin{align}
    {\hat{A}}^{(\infty)}_\hbar(\hat{x},\hat{y})=e^{\hat{x}}{(-1)^{f+1}} q^{1+\frac{f}{2}} e^{{\hat{y}}(1+f)} \prod_{j=1}^r(1-\alpha_j q^{\frac{1}{2}}e^{\hat{y}})-(1-q^{-\frac{1}{2}}e^{\hat{y}})\prod_{j=1}^s(1-\beta_j q^{-\frac{1}{2}}e^{\hat{y}})
\end{align}
with $q=e^\hbar$.
The corresponding solution whose $\hbar$ expansion encodes open GW invariants is given by \eqref{eq:GWfnbpinfty}
\begin{equation}
\label{eq:GWfnbpinftyintro}
\psi^{(\infty)}_f(x) = \sum_{n\geq 0} (-1)^{n(f+1)} q^{\frac{n^2}{2}(1+f) + \frac{n}{2}}
\frac{\prod_{j=1}^r (q^{\frac12}\alpha_j;q)_n}{(q^\frac12;q)_n\prod_{j=1}^s (q^\frac12\beta_j;q)_n} \,e^{nx}
\end{equation}

\item For the choice of basepoint $z_0 =1$, one has the following 
\begin{align}\notag
    {\hat{A}}^{(1)}_{\hbar}(\hat{x},\hat{y})
    =&e^{\hat{x}}(e^{\hat{y}}-1){(-1)^{f+1}} q^{1+\frac{f}{2}}e^{{\hat{y}}(1+f)} \prod_{j=1}^r(1-\alpha_j q^{\frac{1}{2}}e^{\hat{y}})\\
    &-(e^{\hat{y}}-1)(1-q^{-\frac{1}{2}}e^{\hat{y}})\prod_{j=1}^s(1-\beta_jq^{-\frac{1}{2}} e^{\hat{y}})\notag
\end{align}
for which the solution is (for small $e^x$) \eqref{eq:GWfnbp1}
\begin{equation}
\label{eq:GWinvbp1intro}
\psi^{(1)}_f(x) = \sum_{n\geq 0} (-1)^{n(f+1)} q^{\frac{n^2}{2}(1+f) + \frac{n}{2}}
\frac{(q;q)_n\prod_{j=1}^r (q^{\frac12}\alpha_j;q)_n}{(q^\frac12;q)_n (q^2;q)_n\prod_{j=1}^s (q^\frac12\beta_j;q)_n} \,e^{nx}
\end{equation}
\end{itemize}
\subsection{Connection to symmetric quivers} \label{sec:quivers}
Using the following identities, one can now extract the associated symmetric quivers for the wave functions above 
\begin{equation}
\begin{split}
& (\alpha;q)_n = \frac{(\alpha;q)_\infty}{(\alpha q^n;q)_\infty} = \sum_{i,j=0}^\infty (-q^{-1/2} \alpha)^i \alpha^j \frac{q^{i^2/2+nj}}{(q;q)_i (q;q)_j}
\\ & 
\frac{1}{(\beta;q)_n} = \frac{(q^{n}\beta;q)_\infty}{(\beta;q)_\infty} = \sum_{i,j=0}^m (-q^{-1/2}\beta)^i\beta^j\frac{q^{i^2/2+ni}}{(q;q)_i(q;q)_j}.
\end{split}
\end{equation}
Hence we have 
\begin{equation}
\begin{split}
\psi^{(\infty)}_{f} (x) & = \sum_{n\geq 0}\sum_{\substack{l_a\\ a =\{1,...,4\}}}  \sum_{\substack{k_j^{(1)},k_j^{(2)}\\j\in\{1,...,r\}}}\sum_{\substack{k_i^{(1)},k_i^{(2)}\\ i\in\{1,...,s\}}}   
(q^\frac12)^{l_1+2l_2+l_4 + \sum_{i=1}^s k_i^{(2)}+\sum_{j=1}^r  k_j^{(2)}}
\\ & \times 
(-1)^{l_1+l_3} \frac{q^{\frac12 (n^2(1+f)+l_1^2+l_3^2 + \sum_{j=1}^r (k_j^{(1)})^2+\sum_{i=1}^s (k_i^{(1)})^2) +n(l_2+l_3+ \sum_{i=1}^s k_i^{(1)}+\sum_{j=1}^r k_j^{(2)})}}{\prod_{a=1}^4(q;q)_{l_a}\prod_{j=1}^r(q;q)_{k_j^{(1)}} (q;q)_{k_j^{(2)}}\prod_{i=1}^s(q;q)_{k_i^{(1)}}(q;q)_{k_i^{(2)}}}
\\ & \times 
\frac{((-1)^{f+1}q^\frac12 e^x)^n}{(q;q)_n} \prod_{j=1}^r(-\alpha_j)^{k_j^{(1)}} (\alpha_j)^{k_j^{(2)}} \prod_{i=1}^s(-\beta_i)^{k_i^{(1)}} (\beta_i)^{k_i^{(2)}} 
\end{split}
\end{equation}
from which one can read off the quiver. Similarly, for the other choice of basepoint we compute 
\begin{equation}
\begin{split}
\psi^{(1)}_{f} (x) & = \sum_{n\geq 0}\sum_{\substack{l_a\\ a =\{1,...,4\}}}
\sum_{\substack{m_b\\ b =\{1,...,4\}}}\sum_{\substack{k_j^{(1)},k_j^{(2)}\\j\in\{1,...,r\}}}\sum_{\substack{k_i^{(1)},k_i^{(2)}\\ i\in\{1,...,s\}}}  
\\ &
(-1)^{l_1+l_3+m_1+m_3} (q^\frac12)^{l_1+2l_2+l_4 + m_1+2m_2+3m_3+4m_4+\sum_{i=1}^s k_i^{(2)}+\sum_{j=1}^r k_j^{(2)}}
\\ & \times 
 \frac{q^{\frac12 (n^2(1+f)+l_1^2+l_3^2 +m_1^2+m_3^2+ \sum_{j=1}^r (k_j^{(1)})^2+\sum_{i=1}^r (k_i^{(1)})^2) +n(l_2+l_3+ m_2+m_3+\sum_{i=1}^s k_i^{(1)}+\sum_{j=1}^r k_j^{(2)})}}{\prod_{a=1}^4(q;q)_{l_a} \prod_{b=1}^4(q;q)_{m_b} \prod_{j=1}^r (q;q)_{k_j^{(1)}}(q;q)_{k_j^{(2)}}\prod_{i=1}^s (q;q)_{k_i^{(1)}}(q;q)_{k_i^{(2)}}}
\\ & \times 
\frac{((-1)^{f+1}q^\frac12 e^x)^n}{(q;q)_n} \prod_{j=1}^r(-\alpha_j)^{k_j^{(1)}} (\alpha_j)^{k_j^{(2)}} \prod_{i=1}^s(-\beta_i)^{k_i^{(1)}} (\beta_i)^{k_i^{(2)}}
\end{split}
\end{equation}
\footnote{Compared to the expression in equation 4.1 in \cite{Panfil:2018faz}, our partition functions here {\eqref{eq:GWfnbpinftyintro}} and \eqref{eq:GWinvbp1intro} have additional factors of ratios of $q$-Pochhammers. If we compute the DT invariants from them by using 
\begin{equation}
y = \lim_{q\to 1} \frac{\psi_f(qx)}{\psi_f(x)} = (1- x\prod_{j=1}^r{\alpha_j}^{l_j}\prod_{j=1}^s{\beta_j}^{k_j})^{n\Omega_{n,l_1,...,l_r,k_1,...,k_s}},
\end{equation}
we compute slightly shifted results, because of these additional pieces.}
From the expressions of $\psi_f^{(\infty)}$ and $\psi_f^{(1)}$ we see that, different basepoints correspond to different quivers and hence they define different open DT invariants, which can be extracted from using the methods of \cite{Panfil:2018faz}. These two quivers are naturally related by wall-crossing of the open BPS states. We postpone a thorough investigation of this for a future work. Incidentally, we would also like to point that, the $q$-hypergeometric functions \eqref{eq:GWfnbpinfty}, \eqref{eq:GWfnbp1} arising for strip geometries belong to a class of $q$-hypergeometric series, known as Heine's $q$-hypergeometric functions. That the satisfy $q$-difference equations played a crucial role in proving their {\it quantum modularity} in \cite{garoufalidis2022modularqholonomicmodules}. Indeed, consistency of wall-crossing phenomena for 5D BPS states with modularity, led to deeper understanding of mock modular forms \cite{Alexandrov:2016tnf, Alexandrov:2017qhn}. By general consideration from string theory, it is also expected that closed BPS invariants for even compact Calabi-Yau threefolds should give rise to interesting modular functions \cite{Alexandrov:2013mha}. It is natural to expect that this reasoning extends to open BPS invariants as well, and that interesting modular forms should likewise appear in the context of wall-crossing for 3D–5D BPS states. For the examples of strip geometries, it is quite remarkable that they happen to be quantum modular forms.

However the open DT invariants are defined in a different fashion. One definition is through the DT invariants associated with the symmetric quiver \cite{Panfil:2018faz, Ekholm:2018eee}. Another alternative way proposed in \cite{Banerjee:2018syt, Banerjee:2019apt, Banerjee:2024smk,Jeong:2025yys} has its motivation in physics. Consider M-theory on $\mathcal{Y}$ with an M5-brane wrapping $L\times \mathbb{R}^2\times S^1$. Compactifying the M-theory, one obtains a 5D $\mathcal{N}=1$ theory $T^{\mathrm{5D}}[\mathcal{Y}]$ on $\mathbb{R}^4\times S^1$ coupled to a 3d $\mathcal{N}=2$ theory which we call $T^{\mathrm{3D}}[L]$ on $\mathbb{R}^2\times S^1$. Denoting the $\mathbb{R}^2 = \mathbb{R}_x \times \mathbb{R}_t$, where $x,t$ are spatial and temporal directions respectively, a sector of BPS states which is of particular interest are called ``kinky-vortices'' \cite{Banerjee:2018syt}. They are constant in $\mathbb{R}_t$ and evolve along $\mathbb{R}_x$, while attaining constant field configuration at $x\to \pm\infty$. This is a kink solution. In a small patch of the cylinder $S^1\times \mathbb{R}_x$ however they look like vortices. Starting from the data of the mirror curve $A(e^x,e^y) = 0$ and a one form $\lambda = ydx$ obtained by integrating the holomorphic top form on the fiber direction, \cite{Banerjee:2018syt, Banerjee:2019apt} developed techniques for computing BPS degeneracies of these kinky-vortices. \footnote{In the same setup, to obtain a vortex, one needs to consider an M2-brane wrapping a two cycle and the $S^1$, such that $\partial C_2 \cap L$ is a non-trivial one-cycle in $L$. Then in the 3D theory, the BPS configuration corresponds to a point in $\mathbb{R}^2$, extending along $S^1$, amounting to a vortex. A physically motivated open GW/DT correspondence was provided in \cite{Gupta:2024ics}.} 

\subsection{Open BPS states and exponential networks}\label{sec:expnet}
Let us denote by $X=e^x$ and $Y=e^y$ momentarily to emphasize the exponential nature of the variables. 
 Given an algebraic curve $\Sigma$ in $\mathbb{C}^\times_X\times \mathbb{C}^\times_Y$, there is a natural projection $\pi : \Sigma \rightarrow C := \mathbb{C}^\times_X$. However, as was proposed in \cite{Banerjee:2018syt}, it is useful to pass to the infinite cover $\tilde{\pi}: \tilde\Sigma \rightarrow \Sigma$. Thus, we have the covering maps $\tilde{\Sigma} \xrightarrow{\tilde\pi} \Sigma \xrightarrow{\pi} C$. Above $X \in C$, there are infinitely many sheets $(i,N)$ corresponding to points located in $(X,\log Y_i(x) + 2\pi i N)$, with $N\in \mathbb{Z}$. An $\mathcal{E}$-wall is labelled by an ordered pair $(ij,n)$ interpolating between two preimages of $X\in C$, given by $(X,\log Y_i(x) + 2\pi i N)$ and $(X,\log Y_j(x) + 2\pi i (N+n))$, with $n\in \mathbb{Z}$. Working on the $X$-plane, this is achieved by paths crossing branch cuts and logarithmic cuts which were chosen while fixing trivialization. The shape of the wall is given the solution to the differential equation 
\begin{equation}
\label{netweq}
(\log Y_j(X) - \log Y_i(X) + 2\pi i n) \frac{d\log X(t)}{dt} \in e^{i\vartheta} \mathbb{R}_+
\end{equation}
where $t$ is a proper time along the path and for every $\vartheta$ there exists an exponential network $\mathcal{W}(\vartheta)$. \footnote{Exponential networks first introduced in \cite{Eager:2016yxd} and further developed in \cite{Banerjee:2018syt, Banerjee:2019apt,Banerjee:2020moh,Banerjee:2022oed,Banerjee:2023zqc, Banerjee:2024smk} capture BPS spectrum of M-theory on $S^1\times \mathcal{X}$.This amounts to removing the zero-section of $T^\vee C$ and the price to pay is that of working with a non-exact symplectic space $\big((T^\vee C)^\times, dX \wedge \frac{dY}{Y}\big)$ and a multi-valued one form $\lambda = \log Y d\log X$, with logarithmic branch cuts on $\Sigma \rightarrow C$, or parring to the universal cover $\tilde\Sigma$, as was proposed in \cite{Banerjee:2018syt}.}

Each of these $\mathcal{E}$-walls carry also a combinatorial data. To describe it, let us introduce the lattice 
\begin{equation}
\Gamma_{ij,N,N+n} (X) = H_1^{\textrm{rel}}(\tilde\Sigma; (i,N),(j,N+n);\mathbb{Z})
\end{equation}
The paths in relative homology $a\in \tilde\Sigma$ connect one preimage of $X$ labelled by $(i,N)$ with another preimage $(j,N+n)$. We refer to them as soliton path. We also introduce the union of all such charge lattices for different solitons supported on the same wall as $\Gamma_{ij,n}(X) = \amalg_{N\in \mathbb{Z}} \Gamma_{ij,N,N+n} (X)$. The soliton data is an assignment $\mu(a;X)$ for each $a \in \Gamma_{ij,n}(X)$ (often we will drop the second argument $X$). This open 3D-5D BPS state generating functions in this paper pertain to these BPS states. Below, first we briefly recapitulate how to determine these $\mu(a)$ from exponential networks.

Denote by $\nabla^{ab}$ the abelian flat $GL(1)$ connection on $\tilde\Sigma$ and $\nabla^{na}$ the nonabelian $GL(N)$ flat connection on $C$. For an open path $\wp \in C$, associate the parallel transport map 
\begin{equation}
\mathfrak{F}(\wp) = P \exp \int_\wp \nabla^{na}
\end{equation}
%and for open paths $a\in \tilde\Sigma$, the parallel transport 
\begin{equation}
\mathfrak{X}_a = P\exp \int_a \nabla^{ab}.
\end{equation}
There is an algebra of concatenation on the variables $\mathfrak{X}_a$ and one can express $\mathfrak{F}(\wp)$ in terms of $\mathfrak{X}_a$ using detour rules \cite{Banerjee:2018syt,Gaiotto:2012rg}. Flatness of $\nabla^{na}$ implies $\mathfrak{F}(\wp)$ depends only on the homotopy class of $\wp$. This can be leveraged to uniquely fix the data $\{\mu(a)\}$. \footnote{This is one of the steps to construct the nonabelianization map in \cite{Gaiotto:2012rg,Banerjee:2018syt,Banerjee:2024smk}. The second step to determine the 5D indices correspond to computing jumps in this map, corresponding to topology change of the network. An alternative viewpoint to that count was presented in \cite{Eager:2016yxd,Banerjee:2022oed,Banerjee:2023zqc,Banerjee:2024smk}.This computation has in fact origin in the CFIV index and $tt^*$ geometry \cite{Cecotti:1991qv,Cecotti:1992qh,Cecotti:1992rm}, as was explained for 3D-5D case in \cite{Banerjee:2018syt}. An alternative method based supersymmetric localization will be given in \cite{Banerjee:wwp}.} For some examples, the connections of exponential networks with the solutions of $q$-difference equations was investigated in \cite{Grassi:2022zuk,Alim:2022oll,DelMonte:2024dcr}. This subsection is expected to serve as a reminder to the physical aspects of the open BPS states. 

\subsection{$q$-Barnes type integral representation of wave functions} \label{sec:qBarnes}

Here, we write the integral representation for the wave functions presented above. This highlights certain aspects of wall-crossing of the open BPS states, as we explain below concretely in examples. 

Using the identity, rewriting $\frac{1}{(q^\frac12;q)_n} = \frac{(q^{n+\frac12};q)_\infty}{(q^\frac12;q)_\infty}$, one can rewrite $\psi_f^{(\infty)}(x)$ as 
\begin{equation}
\begin{split}
\psi^{(\infty)}_f(x) & = \frac{1}{(q^\frac12;q)_\infty}\sum_{n\geq 0} (-1)^{n(f+1)} q^{\frac{n^2}{2}(1+f) + \frac{n}{2}}
\frac{(q^{n+\frac12};q)_\infty \prod_{j=1}^r (q^{\frac12}\alpha_j;q)_n}{\prod_{j=1}^s (q^\frac12\beta_j;q)_n} \,e^{nx}
\\ & 
= \frac{i(q;q)_\infty^2\prod_{j=1}^r(q^\frac12\alpha_j;q)_\infty}{2\pi(q^\frac12;q)_\infty \prod_{j=1}^s (q^\frac12\beta_j;q)_\infty}
\\ & \times
\int_{c-i\infty}^{c+i\infty} \frac{dz}{z\, (z;q)_\infty} \frac{(q^\frac12/z;q)_\infty\prod_{j=1}^s (q^\frac12\beta_j/z;q)_\infty}{(q/z;q)_\infty \prod_{j=1}^r (q^\frac12\alpha_j/z;q)_\infty}
\exp\bigg[\frac{f}{2}\frac{(\log z)^2}{\log q} - (x-\pi i f)\frac{\log z}{\log q} \bigg]
\end{split}
\end{equation}
where we choose the contour such that $0<|c|<1$, such that simple poles $z=q^{-n}$ contribute for $n\in \mathbb{Z}_{\geq 0}$ with $|z|>0$.
Denoting the integrand as 
\begin{equation}
L_\infty(z) = \frac{1}{(z;q)_\infty} \frac{(q^\frac12/z;q)_\infty\prod_{j=1}^s (q^\frac12\beta_j/z;q)_\infty}{(q/z;q)_\infty \prod_{j=1}^r (q^\frac12\alpha_j/z;q)_\infty},
\end{equation}
one finds that it satisfies a linear $q$-difference equation 
\begin{equation}
L_\infty(qz) = -(z-q^{-\frac12}) z^{r-s} \frac{\prod_{j=1}^s(z-q^{-\frac12}\beta_j)}{\prod_{j=1}^r (z-q^{-\frac12}\alpha_j)} L_\infty(z)
\end{equation}

One can write the other function $\psi_f^{(1)}(x)$ also as a Barnes integral. 
\begin{equation}
\begin{split}
\psi^{(1)}_f(x) & = \sum_{n\geq 0} (-1)^{n(f+1)} q^{\frac{n^2}{2}(1+f) + \frac{n}{2}}
\frac{(q;q)_n\prod_{j=1}^r (q^{\frac12}\alpha_j;q)_n}{(q^\frac12;q)_n (q^2;q)_n\prod_{j=1}^s (q^\frac12\beta_j;q)_n} \,e^{nx}
\\ & 
= \frac{i(q;q)_\infty^3\prod_{j=1}^r(q^\frac12\alpha_j;q)_\infty}{2\pi(q^\frac12;q)_\infty(q^2;q)_\infty \prod_{j=1}^s (q^\frac12\beta_j;q)_\infty}
\\ & \times
\int_{c-i\infty}^{c+i\infty} \frac{dz}{z\, (z;q)_\infty} \frac{(q^\frac12/z;q)_\infty(q^2/z;q)_\infty\prod_{j=1}^s (q^\frac12\beta_j/z;q)_\infty}{(q/z;q)_\infty^2 \prod_{j=1}^r (q^\frac12\alpha_j/z;q)_\infty}
\\ & \qquad\qquad \times
\exp\bigg[\frac{f}{2}\frac{(\log z)^2}{\log q} - (x-\pi i f)\frac{\log z}{\log q} \bigg]
\end{split}
\end{equation}
Denoting the integrand as 
\begin{equation}
L_1(z) = \frac{1}{(z;q)_\infty} \frac{(q^\frac12/z;q)_\infty(q^2/z;q)_\infty\prod_{j=1}^s (q^\frac12\beta_j/z;q)_\infty}{(q/z;q)_\infty^2 \prod_{j=1}^r (q^\frac12\alpha_j/z;q)_\infty}.
\end{equation}
which satisfies 
\begin{equation}
L_1(qz) = \frac{(z-q^{-\frac12})(z-q)}{1-z} z^{r-s} \frac{\prod_{j=1}^s(z-q^{-\frac12}\beta_j)}{\prod_{j=1}^r (z-q^{-\frac12}\alpha_j)} L_1(z)
\end{equation}
Compared to the transformation of $L_\infty(z)$, the second integrand $L_1(z)$ picks up an additional factor under the action of shift operator. This extra ratio indicates the change in the region of $x$-plane where one computes the partition function. 
\begin{remark}\label{remarklinearequation}
    The property that the integrands $L_\infty$ and $L_1$ satisfy a linear $q$-difference equation is reflected in topological recursion by the fact that the dual side (this is Log-TR after changing $x$ and $y$) has no ramification points in $y$, and therefore given explicitly. All dual $\omega_{g,n}^\vee$ have a closed expression \eqref{omgdual} which give rise to a linear $q$-difference equation for the dual wave function $\psi^\vee$ in \eqref{differencey} with basepoint $z_0=\infty$ and \eqref{differenceyy0} in general.
\end{remark}

Picking up different poles by changing the contour, one gets different open DT partition functions. In the toric diagram, this is equivalent to placing the Aganagic-Vafa Lagrangian on different legs. On the mirror side, this is equivalent working in different neighborhoods of $x$. We also notice that each external leg for strip geometry, will correspond to a puncture.  Working around one puncure, this effects transformations on $z$ inside the integrand. To rewrite it in terms of the contour above, one has to append the integrand with extra $z$-dependent Pochhammers. The above two examples correspond to two punctures of $\Sigma$ which are stacked over $X=e^x \to 0$, when projected down to $\mathbb{C}^\times_X$-plane.

\subsection{Example : $\mathbb{C}^3$}

Let us take the example of $\mathbb{C}^3$ for $f=1$. Then the two partition functions become 
\begin{equation}
\begin{split}
\psi_{1,\mathbb{C}^3}^{(\infty)} (x) 
&= \frac{(q;q)_\infty}{(q^\frac12;q)_\infty} \sum_{n\geq 0} \frac{q^{n^2 + \frac{n}{2}} (q^{n+\frac12};q)_\infty }{(q;q)_n (q^{n+1};q)_\infty} e^{nx}
\\ & 
= \frac{i (q;q)_\infty^2}{2\pi (q^\frac12;q)_\infty} \int_{c-i\infty}^{c+i\infty} \frac{dz}{z\,(z;q)_\infty} \frac{(q^\frac12/z;q)_\infty}{(q/z;q)_\infty} \exp\bigg[\frac12 \frac{(\log z)^2}{\log q}-(x-\pi i) \frac{\log z}{\log q}\bigg],
\\ 
\psi_{1,\mathbb{C}^3}^{(1)} (x) 
&= \frac{(q;q)_\infty^2}{(q^\frac12;q)_\infty(q^2;q)_\infty} \sum_{n\geq 0} \frac{q^{n^2 + \frac{n}{2}} (q^{n+\frac12};q)_\infty (q^{n+2};q)_\infty}{(q;q)_n (q^{n+1};q)_\infty^2} e^{nx}
\\ & 
= \frac{i (q;q)_\infty^3}{2\pi (q^\frac12;q)_\infty(q^2;q)_\infty} \int_{c-i\infty}^{c+i\infty} \frac{dz}{z\, (z;q)_\infty} \frac{(q^\frac12/z;q)_\infty(q^2/z;q)_\infty}{(q/z;q)_\infty^2} 
\\ &\qquad \qquad \qquad \qquad\qquad\qquad \times 
\exp\bigg[\frac12 \frac{(\log z)^2}{\log q}-(x-\pi i) \frac{\log z}{\log q}\bigg],
\end{split}
\end{equation}
where we can clearly distinguish the difference in the integrand in both cases. One can rewrite the integrals as $\int_\Gamma dz e^{W(x,z)/\hbar} g(z,\hbar)$ for some decaying function $g(z,\hbar)$ in $z$ (where $q=e^\hbar$). From either of the equations, we read off 
\begin{equation}
W(x,z) = -{\mathrm{Li}}_2(z) + \frac12 (\log z)^2 - (x-\pi i) \log z
\end{equation}
Extremization of this $e^{z\partial W/\partial z} = 1$, and setting $z=e^{-y}$ gives us the curve $e^{x+2y}+e^y - 1 = 0 $.Furthermore, using $\log (t;q)_\infty = -\sum_{n=1}^\infty \frac{1}{n(1-q^n)}$for the two decaying functions one has 
\begin{equation}
\begin{split}
\log g_{1,\mathbb{C}^3}^{(\infty)} (z,q)
& = \frac{1}{\log q}\big({\mathrm{Li}}_2(z) -\frac12 (\log z)^2\big)+\sum_{n=1}^\infty \frac{1}{n(1-q^n)} \big[z^n - q^{n/2}z^{-n}+ q^n z^{-n}\big]
\\ 
\log g_{1,\mathbb{C}^3}^{(1)} (z,q)
& = \frac{1}{\log q}\big({\mathrm{Li}}_2(z) -\frac12 (\log z)^2\big)
\\ & \qquad \qquad\qquad
+\sum_{n=1}^\infty \frac{1}{n(1-q^n)} \big[z^n - q^{n/2}z^{-n}-q^{2n}z^{-n}+2 q^n z^{-n}\big]
\end{split}    
\end{equation}
Clearly, when expanded in $\hbar = \log q$, these two functions have different expansions.
The function $W(x,z)$ defines the critical points. In fact, one can recover the exponential networks from this as ${\mathrm{Im}} \{W(z_+,x) - W(z_-,x)\} = 0$, where $z_\pm = -\frac12 \big\{1-\sqrt{1-4e^x}\big\}$. The middle dimensional contours defining the exponential integrals are based at $z_\pm$. As we move in $x$-plane, the topology of these middle dimensional contours change as much as the functions $g(x,z)$ change. This phenomenon is indicative of wall-crossing of 3D-5D BPS states. \footnote{In terms of exponential networks \cite{Banerjee:2018syt}, moving in the $x$-plane, one finds due to intersections of the $\mathcal{E}$-walls, new trajectories emanate. This signals the existence of new BPS states, responsible for wall-crossing. Note that in this comparison we have kept $\vartheta = \arg (\hbar)$ fixed.}

We also notice the factors that involve the Pochhammers satisfy first order $q$-difference equations 
\begin{equation}
\begin{split}
& F_\infty (z) = \frac{1}{(z;q)_\infty} \frac{(q^\frac12/z;q)_\infty}{(q/z;q)_\infty},\,\, F_\infty (qz) = -(z-q^{-\frac12}) F_\infty (z),
\\ & 
F_1 (z) = \frac{1}{(z;q)_\infty} \frac{(q^\frac12/z;q)_\infty(q^2/z;q)_\infty}{(q/z;q)_\infty^2}, \,\, F_1(qz) = \frac{(z-q^{-1/2})(z-q)}{(1-z)} F_1(z).
\end{split}
\end{equation}
This change in the $q$-difference equation also signifies appearance of new 3D-5D BPS states in the spectrum, as one moves from the neighborhood of one puncture to another. 

We also note the open DT generating series in the quiver form for both the cases. 
\begin{equation}
\psi_{1,\mathbb{C}^3}^{(\infty)} (x)
= \sum_{\substack{n,l,k,\\ m,s}} (-1)^{l+m}
\frac{q^{n^2+\frac{l^2}{2}+\frac{m^2}{2}+n(k+m)}q^{\frac{n}{2}+\frac{l}{2}+k+s}}{(q;q)_n(q;q)_l(q;q)_m(q;q)_k(q;q)_s} e^{nx}
\end{equation}
and 
\begin{equation}
\begin{split}
\psi_{1,\mathbb{C}^3}^{(1)} (x)
& = \sum_{\substack{n,l_1,l_2,k_1,k_2,\\m_1,m_2,s_1,s_2}} (-1)^{l_1+k_1+m_1+s_1}\frac{q^{n^2+l_1^2/2+k_1^2/2+m_1^2/2+s_1^2/2+n(l_2+k_2+m_1+s_1)}}{\prod_{a=1}^2 (q;q)_{l_a}(q;q)_{k_a}(q;q_{m_a}(q;q)_{s_a}}
\\ & \qquad \qquad\times
\frac{e^{nx}}{(q;q)_n}\,q^{n/2+l_1/2+l_2+k_1/2+k_2+m_2/2+3s_1/2+2s_2}
\end{split}
\end{equation}

\subsubsection{Framing $f=-2$}

For this choice of framing, the curve becomes $ e^{2y}-e^y-e^x = 0$ and the quantum curve is given by 
\begin{equation}
\hat{A}_\hbar^{(\infty)}(e^{\hat{x}},e^{\hat{y}}) 
= - e^{\hat{x}}e^{-\hat{y}} - (1-q^{-1/2}e^{\hat{y}}).
\end{equation}
In this case setting $y_0=\infty$, we have $x_0=\infty$ as well. Thus the series solution is of the form 
\begin{equation}
\begin{split}
\psi_{-2,\mathbb{C}^3}^{(\infty)} (x)
&= \sum_{n\geq 0} q^{-n^2-n/2}(q^\frac12;q)_n  e^{-nx}
\\ & 
= \sum_{\substack{n,l,k,\\m,s}} (-1)^{l+m} \frac{ q^{-n^2+l^2/2+m^2/2+nk+ns} q^{-n/2+k/2+m/2+s}e^{-nx}}{(q;q)_n(q;q)_k(q;q)_l(q;q)_m(q;q)_s}
\\ & 
= \frac{i}{2\pi} (q^\frac12;q)_\infty (q;q)_\infty^2 \int_{c-i\infty}^{c+i\infty} \frac{dz}{(z;q)_\infty} \frac{\exp\big[-\frac32 \frac{(\log z)^2}{\log q}+\frac{ (x-\pi i)\log z}{\log q}\big]}{(q^\frac12 /z;q)_\infty(q/z;q)_\infty},
\end{split}
\end{equation}
where $0<|c|<1$. 

On the other hand $y_0 = 0$ implies $e^{x}$ is small. The corresponding $q$-difference equation is given by 
\begin{equation}
\hat{A}_\hbar^{(1)}(e^{\hat{x}},e^{\hat{y}}) 
= - e^{\hat{x}}(e^{\hat{y}}-1)e^{-\hat{y}} - (e^{\hat{y}}-1)(1-q^{-1/2}e^{\hat{y}})
\end{equation}
The solution can be computed as before which is 
\begin{equation}
\begin{split}
\psi^{(1)}_{-2,\mathbb{C}^3}(x) & = \sum_{n\geq 0} (-1)^{n} q^{-\frac{n^2}{2} + \frac{n}{2}}
\frac{(q;q)_n}{(q^\frac12;q)_n (q^2;q)_n} \,e^{nx}
\\ & 
= \sum_{\substack{n,l_1,l_2,k_1,\\k_2,m_1,m_2,\\s_1,s_2}} (-1)^{n+l_1+k_1+m_1+s_1}q^{\frac12(-n^2+l_1^2+k_1^2+m_1^2+s_1^2+n(l_2+k_2+m_1+s_2)}
\\ & \qquad \qquad  \times 
\frac{q^{\frac{n}{2}+\frac{l_1}{2}+l_2+\frac{k_2}{2}+\frac{3m_1}{2}+2m_2+\frac{s_1}{2}+s_2}e^{nx}}{(q;q)_n\prod_{a=1}^2 (q;q)_{l_a}(q;q)_{k_a}(q;q)_{m_a}(q;q)_{s_a}}
\\ & 
= \frac{i (q;q)_\infty^3}{2\pi (q^\frac12;q)_\infty(q^2;q)_\infty} \int_{c-i\infty}^{c+i\infty} \frac{dz}{z\, (z;q)_\infty} \frac{(q^\frac12/z;q)_\infty(q^2/z;q)_\infty}{(q/z;q)_\infty^2} 
\\ &\qquad \qquad \qquad \qquad\qquad\qquad \times 
\exp\bigg[- \frac{(\log z)^2}{\log q}-(x-\pi i) \frac{\log z}{\log q}\bigg]
\end{split}
\end{equation}
In this case the structures of the integrands and quiver generating functions are quite different for $z_0=\infty$ and $z_0=1$ basepoints respectively. This indicates that how one projects the curve to the $x$-plane is quite crucial, As for that matter, in the previous case for $f=1$, the they corresponded to two punctures over $X=0$, whereas here they correspond to one over $X=0$ and the other one over $X=\infty$.

\subsection{Example : Resolved conifold}

The curve becomes 
\begin{equation}
A_{f,\mathrm{con}}(e^x,e^y) = (1-e^y) + (-1)^f e^x e^{(1+f)y} (1-\alpha e^y) = 0.
\end{equation}

First we consider the case with $f=-1$. For the two choices of basepoints, we write down the open DT generating series and the $q$-Barnes integral form. For simplicity, so that the perturbative solution in $e^x$ holds in \eqref{eq:GWfnbpinfty}, we assume $\alpha$ is large for this case. Then one can compute 
\begin{equation}
\begin{split}
\psi_{-1,{\textrm{con}}}^{(\infty)} (x) 
&= \sum_{n\geq 0} \frac{(q^\frac12\alpha;q)_n}{(q^\frac12;q)_n} \, (q^\frac12 e^x)^n
\\ &= 
\sum_{\substack{n,l_1,l_2,\\m_1,m_2,k_1,k_2}} (-1)^{l_1+m_1+k_1}\frac{q^{\frac12(l_1^2+m_1^2+k_1^2+n(l_2+m_2+k_1))}}{{(q;q)_n\prod_{a=1}^2 (q;q)_{l_a}(q;q)_{m_a}(q;q)_{k_a}}} 
\\ & \qquad\qquad\times 
q^{\frac12(n+l_2+m_1+k_1)} \alpha^{l_1+l_2}e^{nx}
\\ & 
= \frac{i(q^\frac12\alpha;q)_\infty (q;q)_\infty^2}{2\pi(q^\frac12;q)_\infty}
\int_{c-i\infty}^{c+i\infty}\frac{dz}{z\,(z;q)_\infty} \frac{(q^\frac12/z;q)_\infty}{(q^\frac12\alpha/z;q)_\infty (q/z;q)_\infty}
\\ & \qquad \qquad \qquad \times 
\exp\bigg[\frac12\frac{(\log z)^2}{\log q}- (x-\pi i)\frac{\log z}{\log q}\bigg],
\end{split}
\end{equation}

\begin{equation}
\begin{split}
 \psi_{-1,{\textrm{con}}}^{(1)} (x) 
&= \sum_{n\geq 0} \frac{(q;q)_n(q^\frac12\alpha;q)_n}{(q^2;q)_n (q^\frac12;q)_n} \, (q^\frac12 e^x)^n 
\\ & =
\sum_{\substack{n,l_1,l_2,m_1,m_2,\\k_1,k_2r_1,r_2,s_1,s_2}} \frac{q^{\frac12(l_1^2+m_1^2+k_1^2+r_1^2+s_1^2+n(l_2+m_2+k_1+r_2+s_1))}}{{(q;q)_n\prod_{a=1}^2 (q;q)_{l_a}(q;q)_{m_a}(q;q)_{k_a}(q;q)_{r_a}(q;q)_{s_a}}} 
\\ & \qquad\qquad\times 
(-1)^{l_1+m_1+k_1+s_1+r_1}q^{\frac12(n+l_2+m_1+k_1+r_1+2r_2+3s_1+4s_2)} \alpha^{l_1+l_2}e^{nx}
\\ & 
= \frac{i(q^\frac12\alpha;q)_\infty (q;q)_\infty^3}{2\pi(q^\frac12;q)_\infty(q^2;q)_\infty}
\int_{c-i\infty}^{c+i\infty}\frac{dz}{z\,(z;q)_\infty} \frac{(q^\frac12/z;q)_\infty(q^2/z;q)_\infty}{(q^\frac12\alpha/z;q)_\infty (q/z;q)_\infty^2}
\\ & \qquad \qquad \qquad \times 
\exp\bigg[\frac12\frac{(\log z)^2}{\log q}- (x-\pi i)\frac{\log z}{\log q}\bigg]
\end{split}
\end{equation}
From the second line of each function, it is clear that the dimension of the quiver matrix changes, as the basepoint changes. Then from the third lines $(0<|c|<1)$, we observe the change in the integrands. In this case, the two punctures in the neighborhood of which the partition functions were computed are $(e^x,e^y) = (1/\alpha,\infty), \, (1,0)$ respectively. Both punctures are of logarithmic type. Using equation \eqref{netweq}, one sees that the trajectories ($\mathcal{E}$-walls) are spirals in the $X=e^x$ coordinate, given by $X = X_*\exp\big[\frac{e^{i\vartheta}t}{2\pi i n}\big]$, $n\in \mathbb{Z}$, for $X_\star = e^{x_\star}$, corresponding the position of the Aganagic-Vafa brane. The stable 3D BPS states are supported between $X_*$ and the logarithmic punctures, and there are two types of them \cite{Grassi:2022zuk, Alim:2022oll}. The solutions above are written in the complements of these spirals on the $\mathbb{C}^\times_X$-plane. Going from one domain of $X$ to another corresponds appearance of new stable 3D BPS states. This is precisely what is captured in the solutions above. 

We next move to the case of framing $f=0$, for which the curve reads as 
\begin{equation}
A_{0,\mathrm{con}}(x,y) = 1-e^y + e^x e^y (1-\alpha e^y).
\end{equation}
Parametrizing $y = \log z$, for basepoint $z_0=\infty$, one has $y_0=\infty$ and for basepoint $z_0=1$, one has $y_0=0$. In both cases $X=e^x$ is small. The quantum curve in the first case is 
\begin{align}
    {\hat{A}}^{(\infty)}_{\hbar,(0,\mathrm{con})}(\hat{x},\hat{y})=-e^{\hat{x}} q e^{{\hat{y}}} (1-\alpha q^{\frac{1}{2}}e^{\hat{y}})-(1-q^{-\frac{1}{2}}e^{\hat{y}})
\end{align}
The solution for this is written in terms of small $e^x$ and is given by 
\begin{equation}
\begin{split}
\psi_{0,{\textrm{con}}}^{(\infty)} (x) 
&= \sum_{n\geq 0} \frac{(q^\frac12\alpha;q)_n}{(q^\frac12;q)_n} \, q^{n^2/2} (q^\frac12 e^x)^n
\\ &= 
\sum_{\substack{n,l_1,l_2,\\m_1,m_2,k_1,k_2}} (-1)^{l_1+m_1+k_1}\frac{q^{\frac12(n^2+l_1^2+m_1^2+k_1^2+n(l_2+m_2+k_1))}}{{(q;q)_n\prod_{a=1}^2 (q;q)_{l_a}(q;q)_{m_a}(q;q)_{k_a}}} 
\\ & \qquad\qquad\times 
q^{\frac12(n+l_2+m_1+k_1)} \alpha^{l_1+l_2}e^{nx}
\\ & 
= \frac{i(q^\frac12\alpha;q)_\infty (q;q)_\infty^2}{2\pi(q^\frac12;q)_\infty}
\int_{c-i\infty}^{c+i\infty}\frac{dz}{z\,(z;q)_\infty} \frac{(q^\frac12/z;q)_\infty}{(q^\frac12\alpha/z;q)_\infty (q/z;q)_\infty}
\\ & \qquad \qquad \qquad \times 
\exp\bigg[\frac{(\log z)^2}{\log q}- (x-\pi i)\frac{\log z}{\log q}\bigg],
\end{split}
\end{equation}

The other choice of basepoint corresponds to the quantum curve 
\begin{align}
    {\hat{A}}^{(1)}_{\hbar,(0,\mathrm{con})}(\hat{x},\hat{y})
    =&-e^{\hat{x}}(e^{\hat{y}}-1) q e^{{\hat{y}}} (1-\alpha q^{\frac{1}{2}}e^{\hat{y}})
    -(e^{\hat{y}}-1)(1-q^{-\frac{1}{2}}e^{\hat{y}})
\end{align}
For the other case, we have
\begin{equation}
\begin{split}
 \psi_{0,{\textrm{con}}}^{(1)} (x) 
&= \sum_{n\geq 0} \frac{(q;q)_n(q^\frac12\alpha;q)_n}{(q^2;q)_n (q^\frac12;q)_n} \,q^{n^2/2} (q^\frac12 e^x)^n 
\\ & =
\sum_{\substack{n,l_1,l_2,m_1,m_2,\\k_1,k_2r_1,r_2,s_1,s_2}} \frac{q^{\frac12(n^2+l_1^2+m_1^2+k_1^2+r_1^2+s_1^2+n(l_2+m_2+k_1+r_2+s_1))}}{{(q;q)_n\prod_{a=1}^2 (q;q)_{l_a}(q;q)_{m_a}(q;q)_{k_a}(q;q)_{r_a}(q;q)_{s_a}}} 
\\ & \qquad\qquad\times 
(-1)^{l_1+m_1+k_1+s_1+r_1}q^{\frac12(n+l_2+m_1+k_1+r_1+2r_2+3s_1+4s_2)} \alpha^{l_1+l_2}e^{nx}
\\ & 
= \frac{i(q^\frac12\alpha;q)_\infty (q;q)_\infty^3}{2\pi(q^\frac12;q)_\infty(q^2;q)_\infty}
\int_{c-i\infty}^{c+i\infty}\frac{dz}{z\,(z;q)_\infty} \frac{(q^\frac12/z;q)_\infty(q^2/z;q)_\infty}{(q^\frac12\alpha/z;q)_\infty (q/z;q)_\infty^2}
\\ & \qquad \qquad \qquad \times 
\exp\bigg[\frac{(\log z)^2}{\log q}- (x-\pi i)\frac{\log z}{\log q}\bigg]
\end{split}
\end{equation}
In this case, the two partition functions correspond to basepoints $(X,Y) = (0,\infty), \, (0,1)$.  Near the first puncture, the one form behaves as $\lambda = \log Y d\log X \sim d(\log X)^2$. Near the second puncture, the one has $\lambda \sim dX$, behaving regularly. Correspondingly, the spectrum of the 3D-5D BPS states are different. Some of their aspects were analyzed using exponential networks in \cite{Banerjee:2019apt}, for this case. 

As in the case of $\mathbb{C}^3$, the integrands transform as (in both cases, as the framing $f$ affects only the exponent non-trivially) 
\begin{equation}
\begin{split}
& G_\infty(z) = \frac{1}{(z;q)_\infty}\frac{(q^\frac12/z;q)_\infty}{(q^\frac12\alpha/z;q)_\infty (q/z;q)_\infty}, \,\, G_\infty(qz) = -\frac{z(z-q^{-\frac12})}{z-q^{-\frac12}\alpha} G_\infty(z) 
\\ & 
G_1(z) = \frac{1}{(z;q)_\infty}\frac{(q^\frac12/z;q)_\infty (q^2/z;q)_\infty}{(q^\frac12\alpha/z;q)_\infty (q/z;q)_\infty^2}, \,\, G_1(qz) = \frac{z}{1-z}\frac{(z-q^{-\frac12})(z-q)}{z-q^{-\frac12}\alpha} G_1(z).
\end{split}
\end{equation}

\begin{remark}
Conversely, starting from a generating series of open DT invariants written in the form of equation \eqref{eq:DTfnintro}, one can write it as the $q$-Barnes integral. Then expanding the integrand and the normalization factors in $\hbar$, where $q=e^\hbar$, term by term integration produces an asymptotic series in $\hbar$. This corresponds to the open GW partition function. Even though they do not depend on $\arg(\hbar)$, because this is only an asymptotic series, it depends on $x$, as we have seen above. This is very close in spirit to the converse of how we implemented $x$--$y$ duality to compute the open DT partition function. \footnote{Moreover these $q$ -Barnes integral expressions are also particularly useful to make statements about quantum modularity of these series, providing a close parallel to the proof of \cite{garoufalidis2022modularqholonomicmodules}. We will investigate this for a larger class of series akin to these in an upcoming work. This will in fact justify the epithet of this cited paper ``modularity can solve effectively a $q$-difference equation''. One more structure can be observed here: starting from a linear $q$-difference, the $q$-Barnes integrals produce directly a solution to a higher order difference equation. We hope that one can utilize this token to reduce the complexity of difference equations of higher degree by going in the opposite direction. These $q$-Barnes representations seem to tie modularity and difference equation at the same root together.}
\end{remark}

\vspace{5 mm}

\bibliographystyle{halpha-abbrv}
\bibliography{omega.bib}
\end{document}